\documentclass{jfm}

\usepackage{etex}
\usepackage{lmodern}
\usepackage{gensymb}
\usepackage[dvips]{graphicx}
\usepackage{url}
\usepackage{setspace}
\usepackage{natbib}
\usepackage[percent]{overpic}
\usepackage{graphics}
\usepackage{amssymb}
\usepackage{amsmath}
\usepackage{mathrsfs}
\usepackage{mathtools}
\usepackage{cancel}
\usepackage{soul}
\usepackage{subfig}
\usepackage{float}
\usepackage{textcomp}
\usepackage{listings}
\usepackage{placeins}
\usepackage{color}
\usepackage{epsfig}
\usepackage{bm,amsbsy}
\usepackage{lscape}
\usepackage[usenames,dvipsnames,table]{xcolor}
\usepackage{tikz}
\usetikzlibrary{positioning,arrows}
\usetikzlibrary{decorations.pathreplacing}
\usepackage{pstricks}
\usepackage[toc,page]{appendix}
\usepackage{upgreek}
\usepackage{caption}
\usepackage{hyperref}
\allowdisplaybreaks

\usepackage{paralist, soul}
\usepackage{paralist}
\usepackage{makecell}
\usepackage{mathrsfs}

\ifCUPmtlplainloaded \else
  \checkfont{eurm10}
  \iffontfound
    \IfFileExists{upmath.sty}
      {\typeout{^^JFound AMS Euler Roman fonts on the system,
                   using the 'upmath' package.^^J}%
       \usepackage{upmath}}
      {\typeout{^^JFound AMS Euler Roman fonts on the system, but you
                   dont seem to have the}%
       \typeout{'upmath' package installed. JFM.cls can take advantage
                 of these fonts,^^Jif you use 'upmath' package.^^J}%
      }
  \else
  \fi
\fi

\ifCUPmtlplainloaded \else
  \checkfont{msam10}
  \iffontfound
    \IfFileExists{amssymb.sty}
      {\typeout{^^JFound AMS Symbol fonts on the system, using the
                'amssymb' package.^^J}%
       \usepackage{amssymb}%
       \let\le=\leqslant  \let\leq=\leqslant
       \let\ge=\geqslant  \let\geq=\geqslant
      }{}
  \fi
\fi

\ifCUPmtlplainloaded \else
  \IfFileExists{amsbsy.sty}
    {\typeout{^^JFound the 'amsbsy' package on the system, using it.^^J}%
     \usepackage{amsbsy}}
    {\providecommand\boldsymbol[1]{\mbox{\boldmath $##1$}}}
\fi

\newcommand\p{\ensuremath{\partial}}

\newcommand\eps{\varepsilon}

\newcommand{\beq}{\begin{equation}}
\newcommand{\eeq}{\end{equation}}
\newcommand{\lagr}{\ensuremath{\mathcal{L}}} 
\newcommand{\uvec}{\ensuremath{\mathbf{u}}} 
\newcommand{\vvec}{\ensuremath{\mathbf{v}}} 
\newcommand{\fvec}{\ensuremath{\mathbf{F}}} 

\renewcommand{\footnoterule}{%
  \kern -3pt
  \hrule width \textwidth height 1pt
  \kern 2pt
}

\newcommand{\EM}[1]{\textcolor{black}{#1}}
\newcommand{\Rone}[1]{\textcolor{black}{#1}} 
\newcommand{\Rtwo}[1]{\textcolor{black}{#1}}  
\newcommand{\Rthree}[1]{\textcolor{black}{#1}}  

\title[]{Designing a minimal baffle to destabilise turbulence in pipe flows}

\author[E. Marensi, Z. Ding, A. P. Willis and R. R. Kerswell]%
{Elena Marensi$^1$\thanks{Email address for correspondence: elena.marensi@ist.ac.at}, Zijing Ding$^2$, Ashley P. Willis$^1$, Rich R. Kerswell$^2$}

\affiliation{$^1$School of Mathematics and Statistics, University of Sheffield, Sheffield S3 7RH, UK\\
$^2$Centre for Mathematical Sciences, University of Cambridge, Cambridge CB3 0WA, UK
}

\pubyear{}
\volume{}
\pagerange{}
\begin{document}

\maketitle

\begin{abstract}
Motivated by the results of recent experiments (K{\"u}hnen {\emph{et al.}}, {\emph{Flow Turb. Combust.}, vol. 100, 2018, pp. 919--943}), we consider the problem of designing a baffle \Rone{(an obstacle to the flow)} to relaminarise turbulence in pipe flows. Modelling the baffle as a spatial distribution of linear drag $\fvec(\mathbf{x},t)=-\chi(\mathbf{x})\mathbf{u}_{tot}(\mathbf{x},t)$ within the flow ($\mathbf{u}_{tot}$ is the total velocity field and \Rtwo{$\chi \ge 0$ a scalar field}), two different optimisation problems are considered \Rtwo{to design $\chi$} at a Reynolds number $Re=3000$. In the first, the smallest baffle defined in terms of a $L_1 $ norm of $\chi$ is sought which minimises the viscous dissipation rate of the flow. In the second, a baffle which minimises the total energy consumption of the flow is treated. Both problems indicate that the baffle should be axisymmetric and radially localised near the pipe wall, but struggle to predict the optimal streamwise extent. A manual search finds an optimal baffle one radius long which is then used to study how the amplitude for relaminarisation varies with $Re$ up to $15\,000$. Large stress reduction is found at the pipe wall, but at the expense of an increased pressure drop across the baffle. Estimates are then made of the break even point downstream of the baffle where the stress reduction at the wall due to the relaminarised flow compensates for the extra drag produced by the baffle.
\end{abstract}

\begin{keywords}
pipe flow, optimisation, turbulence control
\end{keywords}

\section{Introduction}
\label{sec:introduction}

Skin-friction drag associated with turbulent wall flows is the main contributor to energy losses in a wide variety of industrial and technological applications and thus represents a major cause of increase in operating costs and carbon emissions. In the oil and gas industry, for example, the majority of the pumping cost to transport these fluids in pipes is associated with overcoming the frictional drag at the wall boundary \citep{keefe-1998}.
Therefore, any reduction in the turbulent drag,
 or even the complete suppression of turbulence, would have a tremendous societal impact both from an economic and ecological viewpoint.
Recently a novel
 method has been designed which achieves such full relaminarisation by just inserting a stationary obstacle in the core of the pipe in order to flatten the incoming turbulent streamwise velocity profile \citep{kuhnen-etal-2018a}.
Surprisingly, this method was shown in the experiments to completely destabilise turbulence, so that laminar flow was recovered downstream of the baffle.
A first step in modelling the experimental baffle was taken by \cite{marensi-etal-2019}, who theoretically showed the complement of the relaminarisation phenomenon observed in the experiments, that is, the enhanced nonlinear stability of the laminar state due to a flattened base profile.
\Rtwo{Our focus here is to tackle the relaminarisation problem by optimising the design of the baffle to save as much energy as possible.}

\subsection{Flow control}
Many control strategies have been proposed in the past fifty years, both active (an external energy input is needed) and passive (the flow field is manipulated without any supply of energy).
Amongst the active techniques, one of the most popular consists in modifying the near-wall turbulence through large-scale spanwise oscillations
 created either by a movement of the wall or by a body force \citep[see][for a review]{quadrio-2011}. For example, \citet{quadrio-sibilla-2000}
were able to achieve 40\% drag reduction at a friction Reynolds number $Re_{\tau}=172$ by oscillating a pipe around its longitudinal axis, and \citet{auteri-etal-2010} reported a drag reduction of 33\% at $Re_{\tau}\approx 200$ by applying a streamwise-travelling wave of spanwise velocity at the wall.
Passive control strategies include engineered surfaces, e.g. riblets \citep{garcia-jimenez},
 hydrophobic walls \citep{min-kim-2004, aghdam-ricco-2016} and the addition of polymers \citep{toms-1948,virk-etal-1967, owolabi-etal-2017,choueiri-etal-2018}. They have the obvious advantage of requiring no energy input, but in general achieve lower drag reduction than active methods.

Ultimately, the goal of turbulence control is to completely extinguish turbulence, but, in most cases, none of these techniques are able to achieve so. Temporary relaminarisation phenomena have been reported in pipe and channel turbulent flows under the effect of
acceleration,
curvature,
heating,
magnetic field,
and stratification \citep[see][for a review]{sreenivasan-1982}. Interestingly, \citet{he-etal-2016} obtained relaminarisation in a buoyancy-aided flow (vertical pipe heated from below) and showed that the mean flow was flattened by the buoyancy force. The relaminarisation was attributed to the reduction in the ``apparent'' Reynolds number of the flow, only related to the pressure force of the flow.
A flattened base profile is also characteristic of magnetohydrodynamic duct
 flows, for which suppression of turbulent fluctuations is a known phenomenon \citep{krasnov-etal-2008}.

Relaminarisation is not only alluring because of the huge energy savings it would lead to, but is also a very interesting phenomenon from a fundamental point of view as it requires a profound understanding of the mechanisms of production and dissipation of the near-wall turbulence.
 It is well established that in linearly stable flows, such as pipe flow, transition to turbulence occurs via non-modal amplification of small-amplitude cross-flow disturbances to large-amplitude streaks which then break down \citep{schmid-henningson-2001}.
Non-modal growth is associated with the so called lift-up mechanism \citep{brandt-2014} in which the vortices lift low-speed fluid from the wall into the fast moving interior, while the high speed fluid is brought down towards the wall.
This mechanism is also present in fully turbulent flows, where it accounts for the generation of strong velocity streaks induced by the near-wall quasi-streamwise vortices. For turbulence to be self-sustained, though, feedback mechanisms that generate new vortices must also be present, appended to the streak transient growth. These feedback mechanisms have been discussed, amongst others, by \citet{W97}, \citet{jimenez-pinelli-1999} and \citet{schoppa-hussain-2002}, who suggested that the streamwise vortices are regenerated by a secondary instability of the near wall streaks.

Most of the control methods to suppress turbulence have thus focused on targeting different key structures or stages of the turbulence regeneration cycle in order to interrupt it.
For example, \citet{choi-etal-1994} and \citet{xu-etal-2002} developed an opposition control technique aimed at counteracting the streamwise vortices
 by wall transpiration in order to achieve drag reductions, or even a full collapse of turbulence, in Poiseuille flows.
\cite{bewley-etal-2001} used adjoint-variational techniques to derive a model-based optimal control strategy and applied it to wall transpiration in turbulent plane channel flow. 
Another class of feedback control strategies targets the streak-instability vortex regeneration mechanism by eliminating or stabilising the near wall low-speed streaks by means of appropriate spanwise forcing of the flow \citep{du-karniadakis-2000}. 
These methods, although the most sophisticated and advanced on a theoretical basis, are difficult and expensive to implement as they require small scale sensors and actuators
 for real time measurements and control of the flow.
Furthermore, the roll-streak energy growth processappears to be the primary contributor to the turbulent energy production \citep{schoppa-hussain-2002,tuerke-jimenez-2003}, while the precise manner of the turbulent feedback mechanism is secondary.
This suggests that large-scale methods that target the mean shear to counteract/weaken the lift-up mechanism may be the most effective in destroying turbulence. 
 The important role of the
mean shear was confirmed by \citet{hof-etal-2010} in their relaminarisation experiments of localised turbulence.  \Rthree{ At relatively low Reynolds number ($1760 \lesssim Re \lesssim 2300$), turbulence takes the
the form of localised structures, known as puffs, which coexist with the laminar flow 
\citep{WygCha73, WPKM08}.}
 \citet{hof-etal-2010} observed that if two puffs were triggered too close to each other, the downstream puff would collapse. They attributed the relaminarisation of the puff to the flattened streamwise velocity profile induced by the trailing puff.
\Rthree{ A proof-of-concept numerical study was also performed to support this idea. By adding a volume force to the Navier-Stokes equations to flatten the base profile, they were able to suppress turbulence up to $Re = 2900$ (see their supporting material).}
 The flattening indeed reduces the energy supply from the mean flow to the streamwise vortices, thus subduing the turbulence regeneration cycle beyond recovery. 

Recently, there has been a series of experiments \citep{kuhnen-etal-2018a, kuhnen-etal-2018b,kuhnen-etal-2019, scarselli-etal-2019} showing that the flattening of a turbulent streamwise velocity profile in a pipe flow leads to
a full collapse of turbulence
for Reynolds numbers up to $40\,000$, thus reducing the frictional losses by as much as 90\%.
 Different experimental techniques were employed to obtain the flattened base profile -- e.g. rotors or fluid injections to increase the turbulence level near the wall, or an impulsive streamwise shift of a pipe segment to locally accelerate the flow --
 all of them
 being characterised by a reduced linear transient growth, as compared to the uncontrolled case. 
It should be noted that the above-quoted highest Reynolds number reported in the experiments
 was achieved with the wall-movement method,
whose applicability, however, is limited by the fact that the shift length, and hence the time needed to flatten the mean profile, increase linearly with $Re$. \Rthree{In the same spirit as \citet{hof-etal-2010}, \cite{kuhnen-etal-2018a} also performed numerical experiments with a global volume force and showed that the flattening of the base profile could lead to relaminarisation for Reynolds numbers up to $Re=100\,000$.}

The control technique focused upon here is the experimental baffle described by \citet{kuhnen-etal-2018a}.
The baffle decelerates the flow in the middle thereby accelerating it close to the wall (to preserve the volume flux) causing the base profile to be flattened.
 As well as not requiring any energy input, this technique is also extremely simple to implement. With this control scheme, \citet{kuhnen-etal-2018a} were able to completely relaminarise the flow for $Re$ up to 6000 with the friction drag being reduced by a factor of 3.4 sufficiently downstream of the baffle. For very smooth and straight pipes, the authors observed that, once relaminarised, the flow would remain laminar `forever'. For higher $Re$, e.g. $13\,000$, only a temporary relaminarisation could be achieved, but a `local' drag reduction of more than 10\% could still be obtained in a spatially confined region downstream of the device.

\subsection{Flow optimisation}
In pipes and channels, turbulence arises despite the linear stability of the laminar state. The observed transition scenario can thus only be initiated by finite amplitude disturbances \citep[see][for a review]{ESHW07}.
The `smallest' of such disturbances, i.e. the perturbation of lowest energy that can just trigger transition, called the `minimal seed'  \Rthree{\citep{pringle-kerswell-2010}}, provides a measure of the nonlinear stability of the laminar state. It is both of fundamental interest for characterising the basin of attraction of the laminar state, and of practical use, for identifying disturbances that are the `most dangerous', and therefore need avoiding, when turbulence is undesirable.

In the past ten years, variational methods have been successfully used to construct fully nonlinear optimisation problems to find the minimal seeds for transition in different flow configurations
 \citep{pringle-kerswell-2010, pringle-etal-2012,PWK15,monokrousos-etal-2011, rabin-etal-2012, duguet-etal-2013, cherubini-palma-2014, cherubini-etal-2011, cherubini-etal-2012}; see \citet{kerswell-2018} for a review.
In its simplest form, the minimal-seed problem can be stated as follows: among all initial conditions (incompressible \Rthree{and satisfying the boundary conditions}) of a given perturbation energy $E_0$, the optimisation algorithm seeks which disturbance gives rise to the largest energy growth $\mathcal{G}(T,E_0)$ for an asymptotically long time $T$. To find the minimal seed, the initial energy $E_0$ is gradually increased and the variational problem solved until the critical energy $E_c$ is reached where turbulence is just triggered.

From a control point of view, the ability to quantify the nonlinear stability of the laminar state
means that this knowledge can be used to design more nonlinearly stable flows by some manipulation of the system. Indeed, if the critical initial energy for transition of the minimal seed can be shown to increase with some control strategy, then the latter is claimed to be effective. This was the idea underlying the study of \citet{rabin-etal-2014}, where a suitable spanwise oscillation of the wall in Plane Couette flow was shown to increase $E_c$ by 40\%.

Based on the same concept, \cite{marensi-etal-2019} showed enhanced nonlinear stability of a flattened base profile in a pipe, by studying the effect of flattening on the minimal seed. 
Direct numerical simulations (DNS) were also performed using a simple model for the presence of the experimental baffle.
By the no-slip condition, the surfaces of the baffle apply a drag to the flow. Hence, in our simulations \citep{marensi-etal-2019}, we modelled the obstacle as a simple linear drag force of the form $\fvec(\mathbf{x},t)$=$-\chi(\mathbf{x})\mathbf{u}_{tot}(\mathbf{x},t)$, where $\mathbf{u}_{tot}$ is total velocity field and $\chi =\chi(z) \ge 0$ is a step function that introduces a streamwise ($z$) localisation of the baffle
and is homogeneous in the other directions. 
 In \cite{marensi-etal-2019} we showed that turbulence can be avoided by this method, i.e. the basin of attraction of the laminar state is expanded in the presence of the baffle. 

Here, 
 we construct a new fully nonlinear optimisation problem, whereby the minimal baffle, defined in terms of a $L_1$ norm of $\chi(\mathbf{x})$
is sought to just destabilise the turbulence. We allow $\chi(\mathbf{x})$ to be \Rone{any $C^{\infty}$ function of space} and apply an algorithm to find the optimal spatial dependence. This optimisation problem can be viewed as the dual of the minimal-seed problem of \citet{marensi-etal-2019}.

\section{Formulation}
\label{sec:formulation}

\begin{figure}
  \centering
    \includegraphics[width=0.9\textwidth]{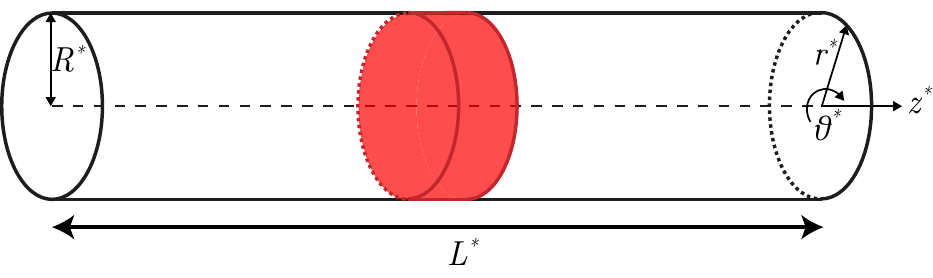}
     \caption{Schematic representation of the pipe geometry with the baffle (in red).}
\label{fig-schematic}
\end{figure}

We consider 
the flow of an \Rone{incompressible Newtonian fluid} 
through a straight cylindrical pipe of radius $R^*$ and length $L^*$ \Rone{(refer to figure \ref{fig-schematic})}, \Rthree{where the symbol $^*$ is used to denote a dimensional quantity}.
The flow is driven by an externally applied pressure gradient, with the mass-flux kept constant, and is subject to a body force 
 which acts against the flow.
The flow is described using cylindrical coordinates $\{r^*,\theta^*,z^*\}$, where $z^*$ is aligned with the pipe axis. 
 Length scales are non-dimensionalised by the pipe radius $R^*$ and velocity components by the unforced centerline velocity $2 U_b^*$, where $U_b^*$ is the constant bulk velocity, \Rthree{so time is in units of $R^*/(2U_b^*)$}. The Reynolds number is $Re:=2 U_b^*\,R^*/\nu^*$, where $\nu^*$ is the kinematic viscosity of the fluid. 
 \Rthree{For the rest of the paper, all quantities will be given in nondimensional units}

The total velocity and pressure fields are decomposed as $\uvec_{tot}=\mathbf{u}_{lam}(r)+\uvec(r,\theta,z,t)$ and $p_{tot}=p_{lam}(z)+p(r,\theta,z,t)$, where $\uvec=\left\{u_r,u_{\theta},u_z\right\}$ and $p$ are the deviations from the unforced laminar velocity field $\mathbf{u}_{lam}:=\mathcal{U}(r)\mathbf{\hat{z}}=(1-r^2)\mathbf{\hat{z}}$ and pressure field $p_{lam}:=-4z/Re$, respectively.
 Following \citet{marensi-etal-2019}, the body force 
 is designed to mimic the drag experienced by the baffle as a linear damping, namely
\beq
 \fvec(\mathbf{x},t):=-\chi(\mathbf{x})\mathbf{u}_{tot}(\mathbf{x},t),
\label{eq:F}
\eeq
 \Rthree{where $\chi(\mathbf{x}):=\phi(\mathbf{x})^2 \ge 0$} 
represents a measure of the level of blockage of the flow by the baffle. The problem is governed by the Navier-Stokes and continuity equations \Rone{for incompressible Newtonian fluid flows}, namely
\beq
\mathbf{NS}:=\frac{\p \uvec}{\p t} + \mathcal{U}\frac{\p \uvec}{\p z} + u_r\,\mathcal{U}'\mathbf{\hat{z}} -\uvec \times \nabla \times \uvec + \nabla p -\frac{1}{Re} \nabla^2 \uvec - \fvec(\mathbf{x},t) =0,
\label{eq:NS}
\eeq
\beq
 \nabla \cdot \uvec=0
\eeq
where the prime indicates total derivative. Periodic boundary conditions are imposed in the streamwise direction and no-slip conditions on the pipe walls.
\Rone{The implementation of the baffle described by \eqref{eq:F} and \eqref{eq:NS}
 is analogous  
to a Brinkman-type penalization technique} \citep{angot-etal-1999}\Rone{, where $\chi$ corresponds to the inverse of the permeability of the porous medium.}
\Rtwo{Improvements to our model, such as dropping the nonlinear terms in the region occupied by the porous baffle} \citep{nield-1991}, \Rtwo{are subjects for future investigation.}

The pressure field is subdivided as $p = \zeta(t)z +\hat{p}(r,\theta,z,t)$
so that $\nabla p = \nabla \hat{p} +\zeta(t) \mathbf{\hat{z}}$, where $\zeta= -4\beta(t)/Re$ is a correction to the pressure gradient such that the mass flux remains constant and $\nabla \hat{p}$ is a (strictly) spatially-periodic pressure gradient. The parameter $1+\beta$ is an observed quantity in experiments and is defined as:
\begin{equation}
\label{eq-1plusbeta}
 1+\beta:= \frac{\langle \p p_{tot} /\p z \rangle_{}}{\langle \p p_{lam} /\p z \rangle} 
\end{equation}
where
the angle brackets indicate the volume integral
\beq
\langle (\bullet) \rangle := \int_0^L \int _0^{2\pi}\int_0^1 (\bullet)\, r\mathrm{d}r\mathrm{d}\theta\mathrm{d}z\,.
\eeq
We also introduce streamwise, azimuthal and cylindrical-surface averages as follows
\beq
\overline{(\bullet)}^z:=\frac{1}{L}\int_0^{L}(\bullet) \,\mathrm{d}z\,, \quad \overline{(\bullet)}^{\theta}:=\frac{1}{2\pi}\int_0^{2\pi}(\bullet) \,\mathrm{d}\theta\,, \quad \overline{(\bullet)}^{\theta,z}:=\frac{1}{2\pi L} \int_0^L\int_0^{2\pi}(\bullet) \,\mathrm{d}\theta \mathrm{d}z\,,
\eeq
as well as a time average 
\beq
\label{eq:timeavg}
\overline{(\bullet)}^{t} := \frac{1}{\tau}\int_{T-\tau}^T (\bullet) \,\mathrm{d}t\,,
\eeq
where $T$ is an asymptotically long time horizon
 and $0<\tau\le T$. 
The force balance in the streamwise direction gives
\beq
\label{eq-NSz}
\beta(t) = \underbrace{- \frac{1}{2}\left.\frac{ \partial \overline{u_{z}}^{\theta,z}}{ \partial r} \right|_{r=1} }_{\mathscr{T}_w} +\underbrace{\frac{Re}{2}  \int_0^1 \overline{-\fvec \cdot \mathbf{\hat{z}}}^{\theta,z} r\mathrm{d}r}_{\mathscr{B}}\,, 
\eeq
where $\mathscr{T}_w$ is the shear stress at the wall and $\mathscr{B}$ is the extra drag due to the local pressure drop across the baffle. In the unforced case the additional pressure fraction $\beta$ has to balance the turbulent wall stress $\mathscr{T}_w$ only.
\Rtwo{By taking the volume integral of the Navier Stokes equations (for the total flow) dotted with $\uvec_{tot}$ we obtain the following energy balance:}
\beq
\label{eq-energy-t}
\frac{1}{2}\left \langle\frac{\partial  \uvec_{tot}^2 }{\partial t} \right \rangle=\underbrace{\langle \uvec_{tot} \cdot (-\nabla p_{tot}) \rangle }_{\mathcal{I}}- \underbrace{\frac{1}{Re} \langle (\nabla \times \uvec_{tot})^2 \rangle}_{ \mathcal{D}} - \underbrace{\langle-\fvec \cdot \uvec_{tot}\rangle}_{\mathcal{W}}\,,
\eeq
\Rtwo{where the term on the left-hand side is the change in kinetic energy and the terms on the right-hand side are, respectively: the input energy $\mathcal{I}(t)= (2 \pi L/ Re) (1+\beta) $  needed to drive the flow, the viscous dissipation $\mathcal{D}(t)$ and the work $\mathcal{W}(t)$  done by the flow against the baffle drag.}
By taking the time average of \eqref{eq-energy-t} we obtain:
\beq
\label{eq-energy}
\overline{\mathcal{I}}^t = \overline{\mathcal{D}}^t + \overline{\mathcal{W}}^t\,.
\eeq
\Rtwo{In the unforced case the input energy has to balance, on average, the viscous dissipation only, i.e. the energy budget reduces to $\overline{\mathcal{I}}^t = \overline{\mathcal{D}}^t$.}

\subsection{The objective functionals}
\label{sec:variational_problem}

The formulation of the variational problem depends on the choice of the objective functional to optimise in order to relaminarise the flow.
The simplest choice is to minimise the total viscous dissipation $\mathcal{D}(\uvec_{tot})$. This method should \Rthree{select a laminar solution if it is stable}. However, as shown by the force balance \eqref{eq-NSz}, the baffle introduces an extra drag, which needs to be taken into account in the overall energy budget \eqref{eq-energy-t}. The key quantity of interest (to be minimised) is thus the total energy input into the flow 
$\mathcal{I}(\uvec_{tot}; \phi)$, which includes the work done $\mathcal{W}(\uvec_{tot};\phi)$ against the baffle.
In either case and in contrast to the minimal seed problem, we need to time average the objective functional. This is done over a time window 
$[(T-\tau),\, T]$ taken to be sufficiently long and sufficiently far from the initial time that the flow can be regarded as statistically steady. \Rone{In fact $\tau=T $ gave the best convergence and was adopted henceforth in the optimisation}.

 Furthermore, to avoid sensitivity to initial conditions, we consider $N>1$ (typically $N=20$ is found to be sufficient) turbulent fields \Rtwo{and perform the optimisation using information from the ensemble of turbulent fields}.
 The two optimisation problems arising from the different choice of objective functional are formulated in the next two sections.
\Rthree{Minimising the total input energy, albeit the most intuitive choice, is expected to be more challenging than minimising the dissipation only, as the algorithm has to simultaneously decrease the dissipation rate and the work done. We therefore introduce the minimisation problem for the viscous dissipation first.} 

\subsection{Optimisation problem 1: design a baffle to minimize viscous dissipation}
\label{subsec:pb1}
The functional to minimise is:
\beq
\mathcal{J}_1 := \sum_n\overline{\mathcal{D}_n}^t(\uvec_{tot,n}) = \sum_n \frac{1}{T}\int_0^T \frac{1}{Re}\langle (\nabla \times \uvec_{tot,n})^2 \rangle \mathrm{d}t\,,
\label{objective_function}
\eeq
where $\overline{\mathcal{D}_n}^t(\uvec_{tot,n}) $ is the time-averaged dissipation associated with the $n^{th}$ turbulent field and $\sum_n$ corresponds to $\sum_{n=1}^N$. 
The above functional is minimised subject to the constraints of the 3D Navier-Stokes equation, constant mass flux and a given amplitude of $\chi(\mathbf{x})$, $\langle \phi^2 \rangle=A_0$.
 Then, motivated by the dual minimal-seed problem, we gradually decrease $A_0$ until we cannot relaminarise the flow any more and thus we have reached the critical (minimal) amplitude $A_{crit}$. The work done associated with a given $\chi$ can be calculated
as an observable following the optimisation. The baffle modifies the mean streamwise velocity profile $U_{mean}(r,t)= (1-r^2)+\overline{u_z}^{\theta,z}$.
Another quantity of interest is the total wall shear stress, relative to the unforced laminar value, namely
\beq
\frac{\mathcal{S}}{\mathcal{S}_{lam}} :=  \left. -\frac{1}{2} \frac{\partial U_{mean}}{\partial r}\right|_{r=1} :=  1 + \mathscr{T}_w(t)\,.
\label{eq-wss} 
\eeq
The Lagrangian is:
\begin{align}
\label{lagrangian-opt1}
\lagr_1 &= \mathcal{J}_1 + \lambda \left[\langle \phi^2(\mathbf{x}) \rangle - A_0\right] +  \sum_n\int_0^T \left \langle \vvec_n \cdot \left[\mathbf{NS} (\uvec_n) + \phi^2(\mathbf{x})\mathbf{u}_{tot,n}(\mathbf{x},t) \right]\right \rangle \mathrm{d}t  \, \\ \noindent \nonumber & + \sum_n\int_0^T \langle \Pi_n \nabla \cdot \uvec_n\rangle \mathrm{d}t + \sum_n\int_0^T\langle \Gamma_n \uvec_n \cdot \hat{\mathbf{z}}\rangle \mathrm{d}t.
\end{align}

In the light of our modelling of the baffle as a linear damping force, the choice of a $L_1$ norm seems the most reasonable, as it measures the amount of baffle material in the pipe.
Appendix \ref{app_A} shows that the results obtained with a $L_2$-normed distribution are similar. Taking variations of $\lagr_1$ and setting them equal to zero we obtain the following set of Euler-Lagrange equations for each turbulent field:\\

\emph{Adjoint Navier-Stokes and continuity equations}
\begin{subequations}
\label{adj-NS}
\begin{align}
\frac{\delta \lagr_1}{\delta \uvec_n} &= \frac{\p \vvec_n}{\p t} + \mathcal{U}\frac{\p \vvec_n}{\p z} -\mathcal{U}'\,v_{z,n} \hat{\mathbf{r}} + \nabla \times (\vvec_n \times \mathbf{u}_n) -  \vvec_n \times \nabla \times \mathbf{u}_n +\nabla \Pi_n \,+ \\ \noindent \nonumber & +\frac{1}{Re} \nabla^2\vvec_n  -\Gamma_n(t)\hat{\mathbf{z}} - \phi^2(\mathbf{x})\vvec_n +\frac{2}{Re \, T}\nabla^2\uvec_{tot,n} =0\,,\\
 \frac{\delta \lagr_1}{\delta p_n} &=  \nabla \cdot \vvec_n=0\,.
\end{align}
\end{subequations}

\emph{Compatibility condition} (terminal condition for backward integration)
\beq
\frac{\delta \lagr_1}{\delta \uvec_n(\mathbf{x},T)}=\vvec_n(\mathbf{x},T)= 0\,.
\label{compat}
\eeq

\emph{Optimality condition}
\beq
\frac{\delta\lagr_1}{\delta \phi} = \phi\left(\lambda + \sigma\right) \EM{= 0}\,,
\label{optim}
\eeq

where 
\beq
\label{eq-sigma}
\sigma(\mathbf{x}) := \sum_n\int_0^T\mathbf{u}_{tot,n}\cdot \vvec_n \mathrm{d}t
\eeq
 is a scalar function of space. 
\EM{As the optimality condition \eqref{optim} is not satisfied automatically, $\phi$ is moved in the descent direction of $\lagr_1$ to make the latter approach a minimum where $\delta\lagr_1/\delta \phi$ should vanish.}
The minimisation problem is solved numerically using an iterative algorithm similar to that adopted in \cite{pringle-etal-2012} (see their section 2).
 The update for $\phi$ at the $(j+1)^{\text{th}}$ iteration is
\beq
\phi^{(j+1)}=\phi^{(j)} - \epsilon  \frac{\delta\lagr_1}{\delta \phi^{(j)}} = \phi^{(j)} -\epsilon \phi^{(j)}\left(\lambda + \sigma^{(j)}\right).
\label{update}
\eeq
To find $\lambda$ we impose that the updated $\phi$ satisfies the amplitude constraint, namely
\beq
\left \langle \left[ \phi^{(j+1)}\right]^2(\mathbf{x}) \right \rangle =A_0 \implies \left \langle \left[ \phi^{(j)}\left(1-\epsilon \sigma^{(j)}\right)- \phi^{(j)} \epsilon \lambda  \right]^2 \right \rangle =A_0.
\label{normalisation}
\eeq
The same strategy as \cite{pringle-etal-2012} is employed for the adaptive selection of $\epsilon$. 
Due to the factor $\phi$ in front of the bracket in \eqref{optim}, and thus in \eqref{update},
the choice $\chi=\phi^2$ 
(or $\phi$ to any power greater than 1)
 prevents $\phi$ from becoming non zero in regions of the domain where it was initially zero
(i.e. if $\phi$ is zero somewhere, it cannot change). This issue is overcome by ensuring that the algorithm is fed with an initial guess for $\phi$ which is strictly positive everywhere in the domain, as prescribed in \S \ref{subsec:res_IG}.

\subsection{Optimisation problem 2: design a baffle to minimise the total energy input}
\label{subsec:pb2}
The functional to minimise is:
\beq
\mathcal{J}_2 := \sum_n\overline{\mathcal{I}_n}^t(\uvec_{tot}; \phi)=\sum_n\underbrace{\frac{1}{T}\int_0^T \frac{1}{Re}\langle (\nabla \times \uvec_{tot,n})^2 \rangle \mathrm{d}t}_{\overline{\mathcal{D}_n}^t} + \underbrace{\frac{1}{T}\int_0^T \langle \phi^2\uvec_{tot,n}^2 \rangle \mathrm{d}t}_{\overline{\mathcal{W}_n}^t}\,,
\label{objective_function-2}
\eeq
where $\overline{\mathcal{I}_n}^t$, $\overline{\mathcal{D}_n}^t$ and $\overline{\mathcal{W}_n}^t$ are the time-averaged energy input, dissipation and work done, respectively, associated with the $n^{th}$ turbulent field. The above functional is minimised subject to the constraints of the 3D Navier-Stokes equation and constant mass flux. With this choice of objective functional, we do not need a constraint on the amplitude of $\chi$, because the latter appears in the definition of $\mathcal{W}$ (the work done can be regarded as proportional to $\langle \chi \rangle = \langle \phi^2 \rangle =A_0$), and thus of $\mathcal{J}_2$. 
 The Lagrangian for this problem is
\begin{align}
\label{lagrangian-opt2}
\lagr_2 &= \mathcal{J}_2 + \sum_n\int_0^T \left \langle \vvec_n \cdot \left[\mathbf{NS} (\uvec_n) + \phi^2(\mathbf{x})\mathbf{u}_{tot,n}(\mathbf{x},t) \right]\right \rangle \mathrm{d}t \, +  \\ \noindent \nonumber & +  \sum_n\int_0^T \langle \Pi_n \nabla \cdot \uvec_n\rangle \mathrm{d}t  + \sum_n\int_0^T\langle \Gamma_n \uvec_n \cdot \hat{\mathbf{z}}\rangle \mathrm{d}t
\end{align}
and details of the formulation are given in appendix \ref{app_B}. Note that \eqref{lagrangian-opt2} is the same as \eqref{lagrangian-opt1} with $\mathcal{J}_1$ replaced by $\mathcal{J}_2$  and with the amplitude constraint dropped.\\

A spectral filtering is also applied to \Rone{smoothen $\chi(\mathbf{x})$.}
The formulations for both optimisation problems with spectral filtering are reported in appendix \ref{app_C} for completeness.

\subsection{Numerics}
\label{sec:numerics}
The calculations are carried out using the open source code \verb+openpipeflow.org+ \citep{willis-2017}. 
\Rone{At each time step, the unknown variables, i.e. the velocity and pressure fluctuations $\{\mathbf{u}, p\}$}
are discretised in the domain $\{r,\theta, z\}=[0,1]\times[0,2\pi]\times[0,2\pi/k_0]$, where $k_0=2\pi/L$, using Fourier decomposition in the azimuthal and streamwise direction and finite difference in the radial direction, i.e.
\beq
\{\mathbf{u}, p\}(r_s,\theta,z) = \sum_{k<|K|} \sum_{m<|M|} \{\mathbf{u}, p\}_{s\,k\,m}e^{i k_0 k z + m \theta}\,,
\eeq
where $s=1,...,S$ and the radial points are clustered close to the wall.
\Rthree{Temporal discretisation is via a second-order predictor-corrector scheme, with Euler predictor for the nonlinear terms and Crank-Nicolson corrector.}
The optimisation is carried out at a Reynolds number $Re=3000$ for which 
\Rthree{turbulence is sustained in the absence of control \citep{BSMLAH15}}
 and the computational cost of the iterative algorithm still manageable.

We consider two cases: the first has similar parameters to \cite{pringle-etal-2012}, i.e. $L=10$, $T=300$ (preliminary tests were carried out to verify that the chosen target time is sufficiently long), while the second uses a long pipe $L=50$ (in order to encourage any streamwise localisation in $\chi$) with $T=100$. In the latter time horizon,
 the flow passes through the obstacle only once for the chosen pipe length.
In this way we expect to help break the 
\Rone{axial homogeneity}
 of $\sigma$ (defined in \eqref{eq-sigma}), or of $\tilde{\sigma}$ (defined in \eqref{sigma-tilde}), due to the translational symmetry of the Navier-Stokes and the adjoint equations, which makes the algorithm move towards a fairly streamwise homogeneous $\chi(\mathbf{x})$, as we shall discuss later. In the $L=10$ case, we use $S=60$, $M=32$, $K=48$, while for the long-pipe case we use $K=192$ (and same $S$ and $M$). In both cases the size of the time step is $\Delta t=0.01$.

We also performed DNS at Reynolds number up to $15\,000$ for $L=50$, with the spatial discretisations appropriately increased (e.g. $S=128$, $M=128$, $K=768$ for the largest Reynolds number considered) \Rthree{to ensure a drop in the energy spectra by at least 4 orders of magnitude. In addition, for  $Re \ge 5000$ the time-step size is dynamically controlled using information from the predictor-corrector scheme \citep{willis-2017} and is typically around $0.005$ at $Re=15\,000$}.
\section{Baffle design}
\label{sec:results}
\subsection{Optimisation problem 1 with $N=1$ turbulent field}
\label{subsec:res_IG}
\begin{figure}
  \centering
    \includegraphics[width=1\textwidth]{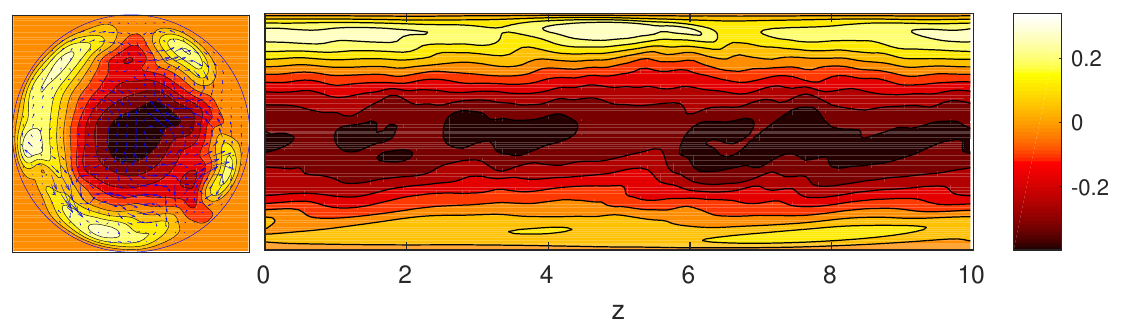}
     \caption{Typical turbulent field used as initial condition in our optimisation algorithm \Rthree{(either for optimisation problem 1 or 2)} at $Re=3000$ and $L=10$. Cross sections: (left) in the $r-\theta$ plane at $z=0$ and (right) in the $r-z$ plane (not in scale) at $\theta=0$. The contours indicate the streamwise velocity perturbation while the arrows in the $r-\theta$ plane correspond to cross-sectional velocities.}
\label{fig-optforcic-ic}
\end{figure}

\begin{figure}
  \centering
    \includegraphics[width=0.48\textwidth]{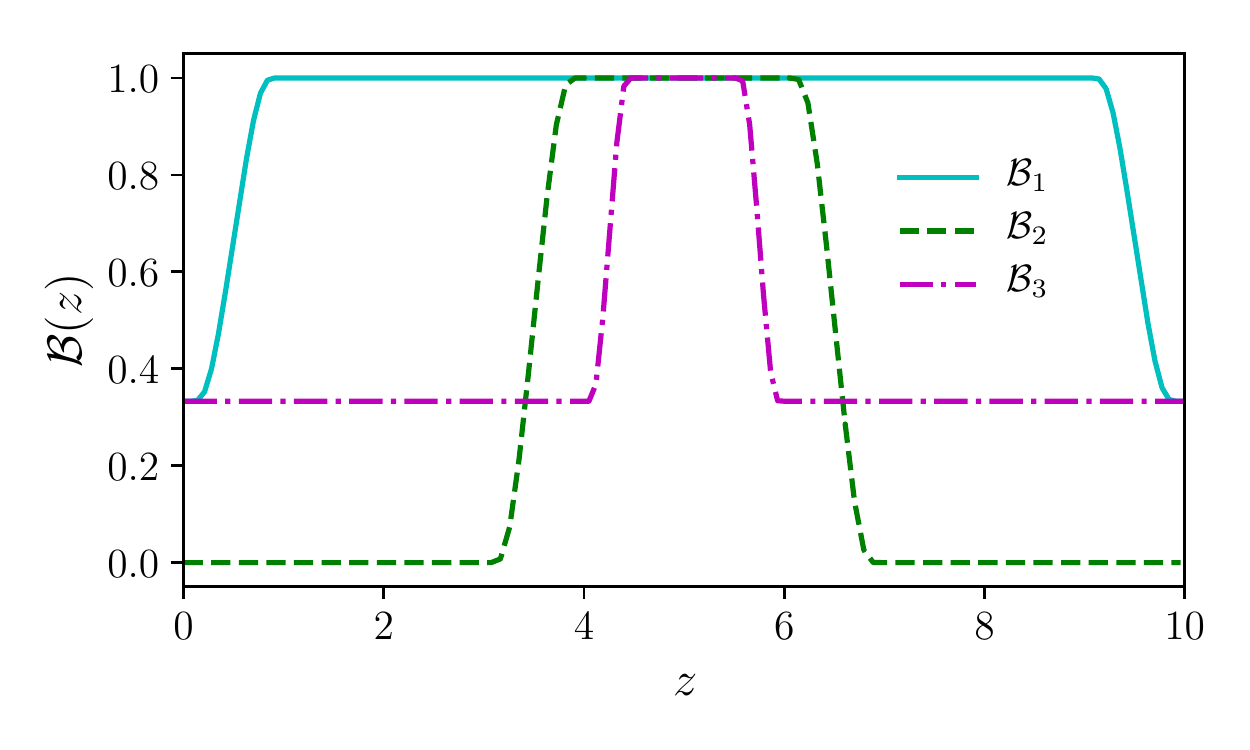}
    \includegraphics[width=0.48\textwidth]{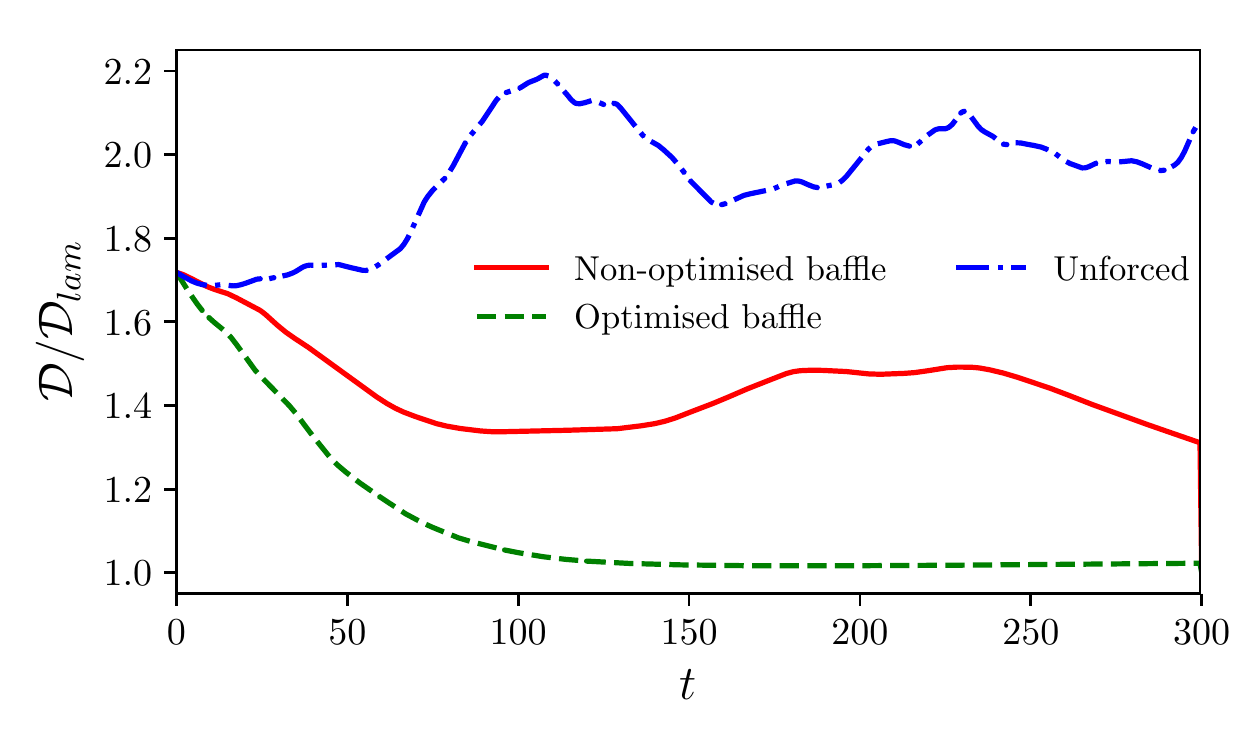}
     \caption{Left: Different \Rthree{streamwise} modulations $\mathcal{B}_i(z)$, $i=1,2,3$ of $\chi$ used as initial guesses in a $L=10$ pipe. The \Rthree{streamwise} modulation $\mathcal{B}_2$ corresponds to the non-optimised baffle of \citet{marensi-etal-2019}. Right: Effect of the non-optimised (with \Rthree{streamwise} modulation $\mathcal{B}_2$) and the optimised baffle (obtained with the latter as initial guess) on a typical turbulent field (shown in figure \ref{fig-optforcic-ic}) at $Re=3000$.}
\label{fig-cfr-smoothing-2}
\end{figure}

Our optimisation algorithm was first tested for the case with $N=1$ turbulent initial field. This study provided suitable initial $\phi$ for the computationally far more expensive case $N=20$. A typical turbulent initial condition in a $L=10$ pipe at $Re=3000$ is shown in figure \ref{fig-optforcic-ic}.
Following \citet{marensi-etal-2019} (refer to their equation 3.5),
we start with the initial guess for $\phi$,
\beq
 \phi^2 = \chi(z) = \mathcal{A}\, \mathcal{B}(z)\,,
\eeq
where $\mathcal{A}$ is a scalar constant to adjust the amplitude of $\chi$ and $\mathcal{B}(z)$ is a (scalar) smoothed step-like function that introduces a streamwise localisation of the force. In \citet{marensi-etal-2019} the smoothing function was defined as \cite[][equation 8]{yudhistira-skote-2011}:
\beq
\mathcal{B}(z)=g\left(\frac{z-z_{start}}{\Delta z_{rise}}\right) - g\left(\frac{z-z_{end}}{\Delta z_{fall}}+1\right)\,,
\label{smoothing-function}
\eeq
with
\[
 g(z^{\dagger}) =
  \begin{cases}
   0 & \text{if } z^{\dagger} \leq 0 \\
   \left\{1+\exp[1/(z^{\dagger}-1)+1/z^{\dagger}]\right\}^{-1} & \text{if } 0<z^{\dagger}<1 \\

   1       & \text{if } z^{\dagger} \geq 1
  \end{cases}
\]
where $z_{start}$ and $z_{end}$ indicate the spatial extent over which $\fvec$ is non-zero and $\Delta z_{rise}$ and $\Delta z_{fall}$ are the rise and fall distances. By construction, $0\le\mathcal{B}(z)\le 1$ $\forall z$.
As explained in \S \ref{subsec:pb1}, to make sure that the initial guess for $\chi$ is strictly positive everywhere, we redefine \eqref{smoothing-function} as follows: 
$\widetilde{\mathcal{B}}= (1-b)\mathcal{B} + b$, with $b\ge 0$ so that $b\le\widetilde{\mathcal{B}} (z)\le 1$ $\forall z$.
 Unless otherwise specified, 
 we use $b=1/3$, so that the initial guess for $\chi$ goes to a third at the sides instead of going to zero, and the tilde will be dropped in the ensuing discussion.
\EM{Using \eqref{smoothing-function} we define the baffle length $L_b$ as the region where $\chi$ attains its maximal values (i.e. where $\mathcal{B}(z)=1$), namely 
\beq
L_b:=(z_{end} -z_{start}) - \Delta z_{rise} - \Delta z_{fall}\,.
\label{eq:Lb}
\eeq
The baffle used in \citet{marensi-etal-2019} (indicated with $\mathcal{B}_2$ in table \ref{table} and figure \ref{fig-cfr-smoothing-2}(left)) was chosen so that it
occupies a fifth of a $L=10$ pipe.}

\begin{table}
\setlength{\tabcolsep}{10pt}
\renewcommand{\arraystretch}{1.5}
\centering
\begin{tabular}[t]{ c|cccccc } 
 Streamwise modulation & $z_{start}$ & $z_{end}$ & $\Delta z_{rise}$ &$\Delta z_{fall}$ &  $b$ &$L_b$\\
\hline
 $\mathcal{B}_1(z)$ & 0 & 10 &1   &1    &1/3 &8 \\ 
 $\mathcal{B}_2(z)$ & 3 & 7  &1   &1    &0 &2  \\  
 $\mathcal{B}_3(z)$ & 4 & 6  &0.5 &0.5  &1/3 &1 \\
\end{tabular}
\caption{\EM{Summary of the parameters used in \eqref{smoothing-function} to characterise the different streamwise modulations $\mathcal{B}_i(z)$, $i=1,2,3$ of the baffle in a $L=10$ pipe, see figure \ref{fig-cfr-smoothing-2}(left). The streamwise modulation $\mathcal{B}_2$ corresponds to the non-optimised baffle of \citet{marensi-etal-2019}. The last column reports the corresponding baffle extent $L_b$, defined in \eqref{eq:Lb}.} \label{table}}
\end{table} 

To check how well our optimisation algorithm performs compared to the available data (the non-optimised baffle), we perform the optimisation starting from the same \Rthree{streamwise} modulation used in \citet{marensi-etal-2019}. Figure \ref{fig-cfr-smoothing-2}(right) shows that the turbulent trajectory is fully relaminarised by the optimised baffle, while it was only `weakened' by the non-optimised baffle at the same $\left \langle \chi \right \rangle =\left \langle \phi^2 \right \rangle = A_0$.

\begin{figure}
  \centering
    \includegraphics[width=0.48\textwidth]{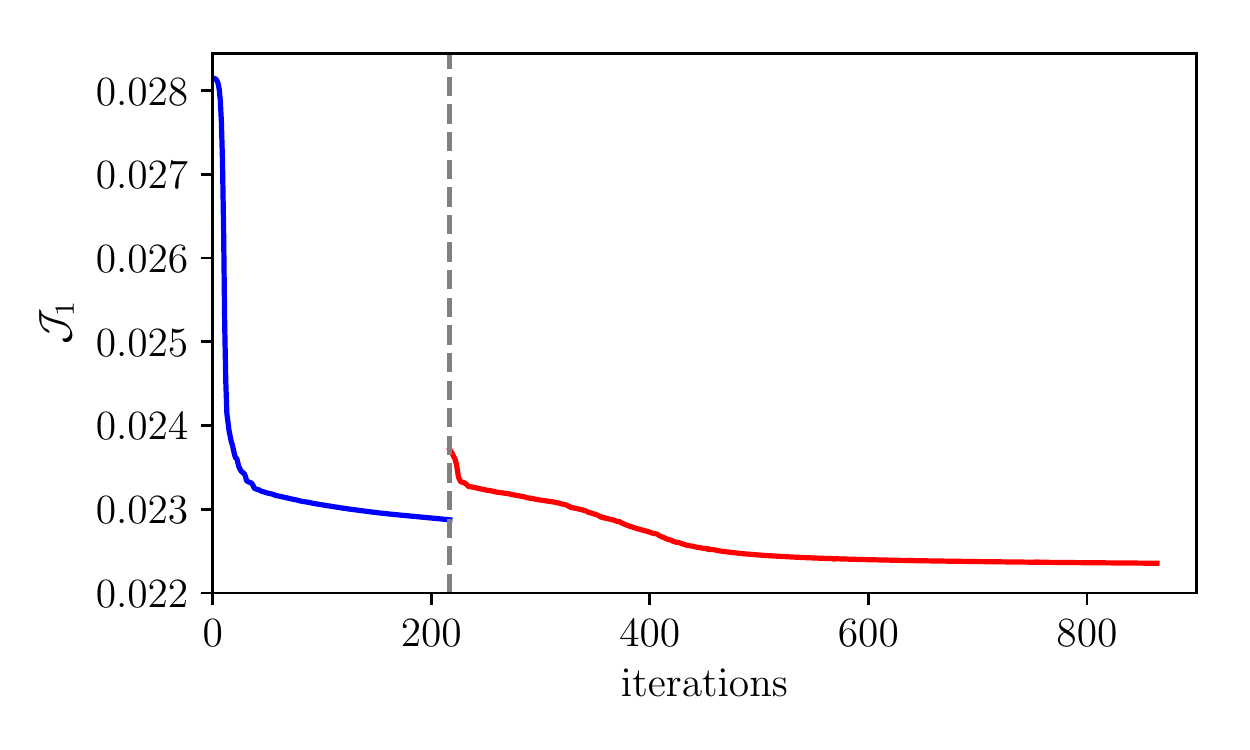}
    \includegraphics[width=0.48\textwidth]{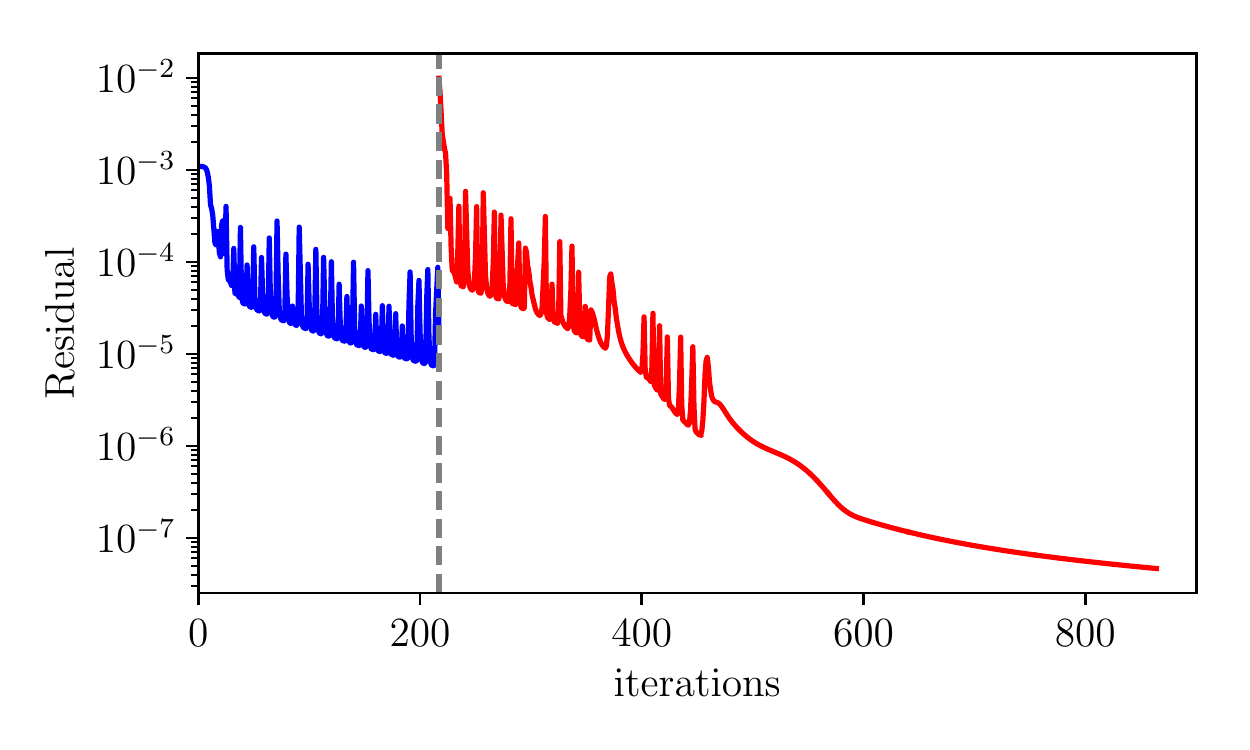}
     \caption{\Rone{Convergence of the algorithm as the code iterates for the case with initial guess $\mathcal{B}_3(z)$ and $N=1$. Left: Objective function $\mathcal{J}_1= \overline{\mathcal{D}}^t$. Right: residual $\left \langle (\delta \mathcal{L}_1/\delta \phi)^2\right \rangle$.} The jump at iteration 217 is due to the algorithm being restarted with different parameters (different $A_{0}$ and a spectral filtering applied in the azimuthal direction \Rthree{to retain only the $m=0$ mode}) to aid convergence.}
\label{fig-cgce}
\end{figure}

 Different \Rthree{streamwise} modulations have been tested as initial guesses for $\phi$. However, in presenting the results, we will focus on two cases, $\mathcal{B}_1(z)$ and  $\mathcal{B}_3(z)$ (see table \ref{table} and figure \ref{fig-cfr-smoothing-2}(left)), which correspond to a wide and a thin baffle, respectively. 
 These two very different initial guesses are considered the most relevant to illustrate the outcomes of our optimisation.
As in \cite{pringle-etal-2012}, the algorithm was checked for convergence by monitoring the residual $\left \langle (\delta \mathcal{L}_1/\delta \phi)^2\right \rangle $ (see \eqref{optim}) and the objective function $\mathcal{J}_1= \overline{\mathcal{D}}^t$ (see \eqref{objective_function}), as the code iterates. A typical example of a converged optimisation is shown in figure \ref{fig-cgce}. The residual has dropped by five orders of magnitude (below $10^{-7}$) and the dissipation has reached a plateau. Note that the jump after approximately 200 iterations is due to the algorithm being restarted with different parameters (different $A_{0}$ and a spectral filtering applied in the azimuthal direction \Rthree{to retain only the $m=0$ mode}) to aid convergence. 
\Rone{It should be pointed out that, while a run with $N=1$ could be very well converged, the resulting optimised $\chi$ may not be able to relaminarise a different turbulent initial condition (hence the necessity of considering $N>1$). 
This $\chi$, however, provides a very good initial guess
for a more expensive optimisation with $N=20$}

Using the $L_1$-norm to measure the baffle amplitude provides another useful check on convergence. For the optimality condition \eqref{optim} to be satisfied at a given spatial location in the flow either: i) $\phi$ vanishes (so the baffle is absent there); or ii) $\lambda+\sigma(\boldsymbol{x})$ vanishes;
 or iii) both. The fact that relaxing the constraint $\chi=\phi^2 \geq 0$ to $\chi=\phi^2-1 \geq -1$ at any point cannot {\em increase} the minimum $\lagr$ (the set of allowable fields is only increasing) means that
\beq
\frac{\delta \lagr}{\delta \phi^2}=\lambda + \sigma(\boldsymbol{x}) \geq 0
\eeq 
at the minimum so then
\beq
\lambda=\max_{\boldsymbol{x}} (-\sigma(\boldsymbol{x})\,)
\eeq
there. As a result $-\sigma(\boldsymbol{x})/\lambda \leq 1$ everywhere at convergence with strict equality necessary (but not sufficient) at points where the baffle is present ($\chi >0$). If $\max_{\boldsymbol{x}} (-\sigma(\boldsymbol{x})\,)$  occurs at isolated points, the baffle takes on the form of a series of $\delta$ functions (see appendix \ref{app_D}). In contrast, if the set of $\boldsymbol{x}$  which maximize  $-\sigma(\boldsymbol{x})$ form a connected domain, the optimal baffle might be degenerate with different optimal baffles having different subsets of support within the domain (see appendix \ref{app_D}). Figure \ref{fig-sigoverlam} shows the tendency of the algorithm towards this  latter situation, with a connected (quasi streamwise-homogeneous) region close to the wall where $-\sigma/\lambda = 1$ (corresponding to where the baffle, i.e. $\chi$, concentrates, as we shall see later) and only small pockets of the domain where $1< -\sigma/\lambda \lesssim 2$ (corresponding to where the baffle is small) indicating convergence is still not complete (initially $-\sigma/\lambda \approx 10$ in places).

\begin{figure}
  \centering
    \includegraphics[width=0.48\textwidth]{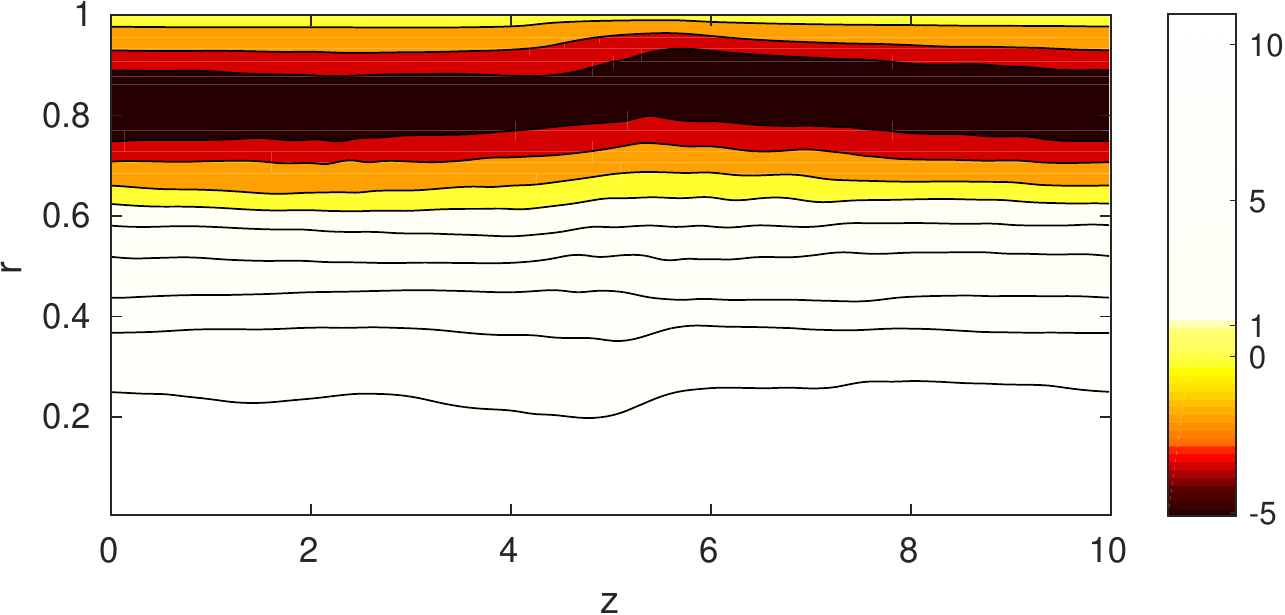}
    \includegraphics[width=0.48\textwidth]{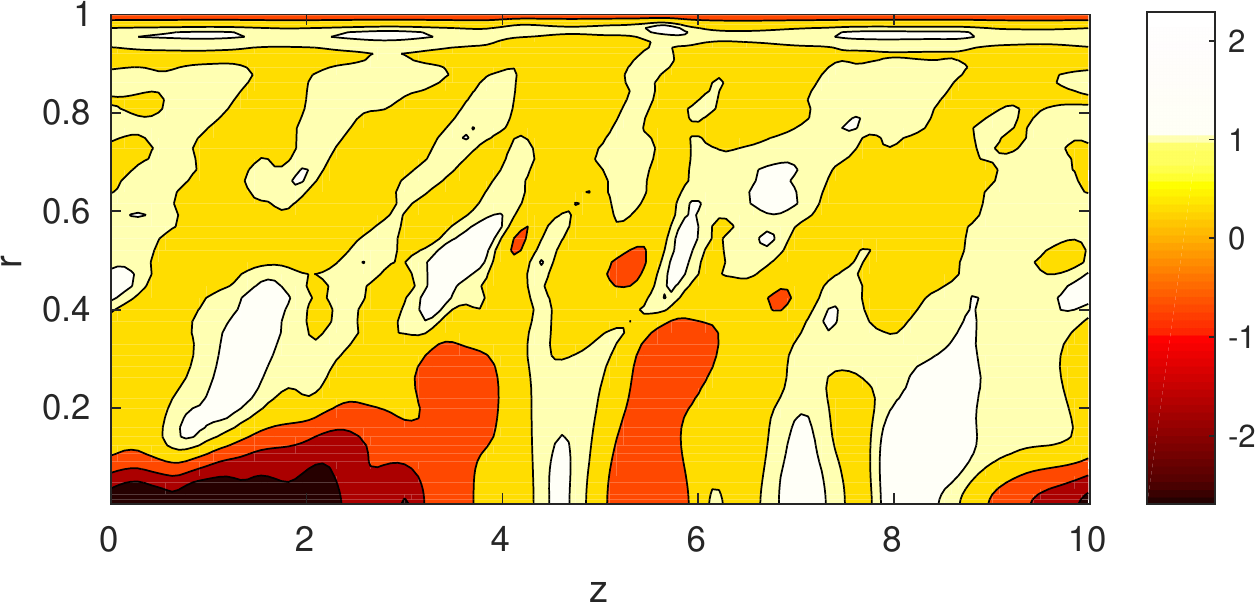}
     \caption{Cross sections of $-\sigma/\lambda$ at the first (left) and last (right) iteration of the case shown in figure \ref{fig-cgce}. The colormap is scaled so that regions where $-\sigma/\lambda>1$ appear in white.}
\label{fig-sigoverlam}
\end{figure}

These preliminary runs showed that $\chi$ tends to be fairly axisymmetric, as expected, given the geometry of the problem. Therefore we apply a spectral filter to filter out $m>0$ modes. All the results presented hereinafter pertain to the case of an axisymmetric baffle $\chi=\chi(r,z)$. 

\subsection{Optimisation problem 1 with $N=20$ turbulent fields}
\label{sec:result_pb1}

\begin{figure}
  \centering
    \includegraphics[width=0.48\textwidth]{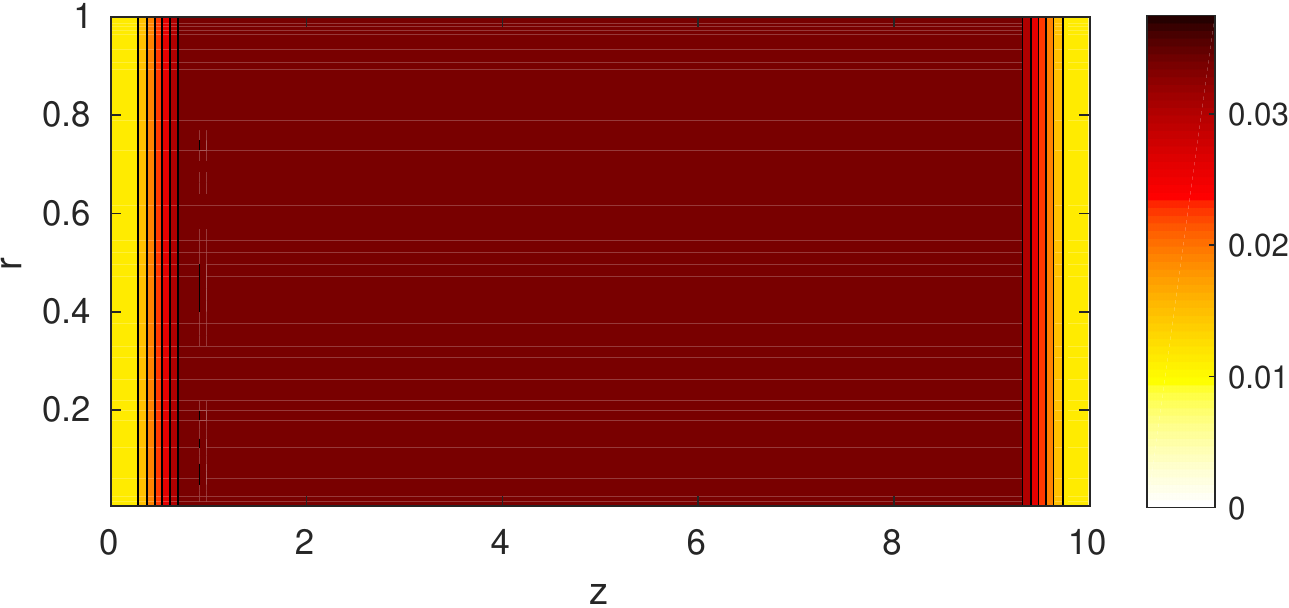}\quad 
    \includegraphics[width=0.48\textwidth]{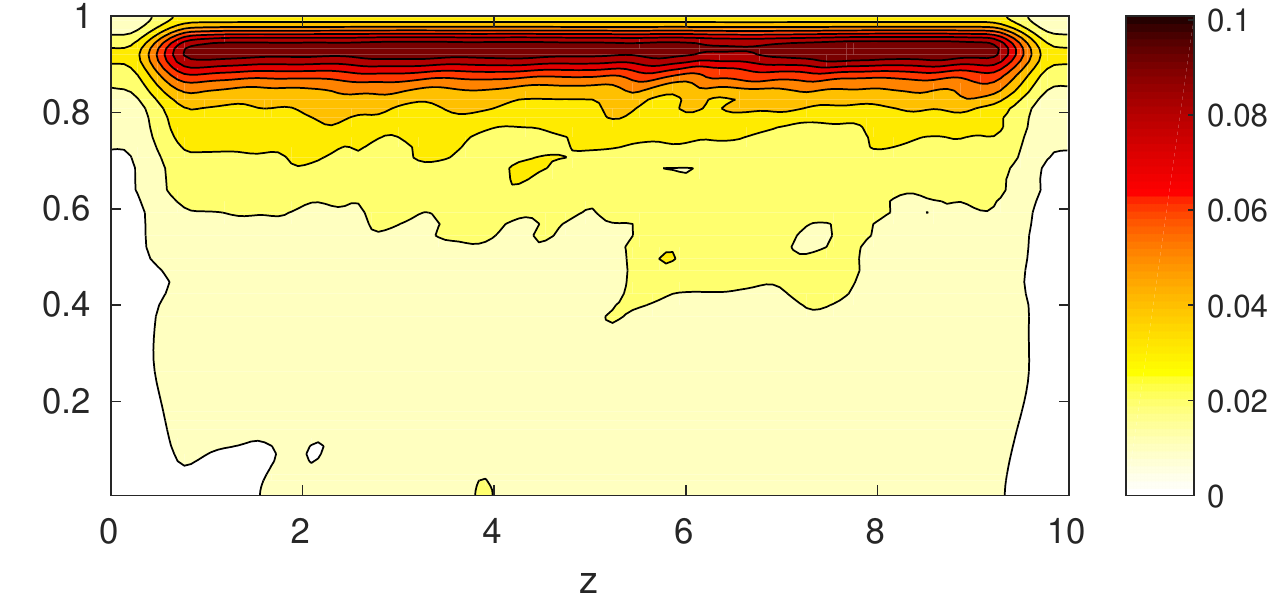}
     \caption{Cross sections in the $r-z$ plane of (left) initial guess for $\chi$ and (right) the converged $\chi$. Ten levels are used between zero and the maximum value of $\chi$. Case $L=10$, $T=300$, $\mathcal{B}_1$ as initial guess. The initial amplitude of $\chi$ is $A_0=1.1$.} 
\label{fig-verywide}
\end{figure}

\begin{figure}
  \centering
    \includegraphics[width=0.48\textwidth]{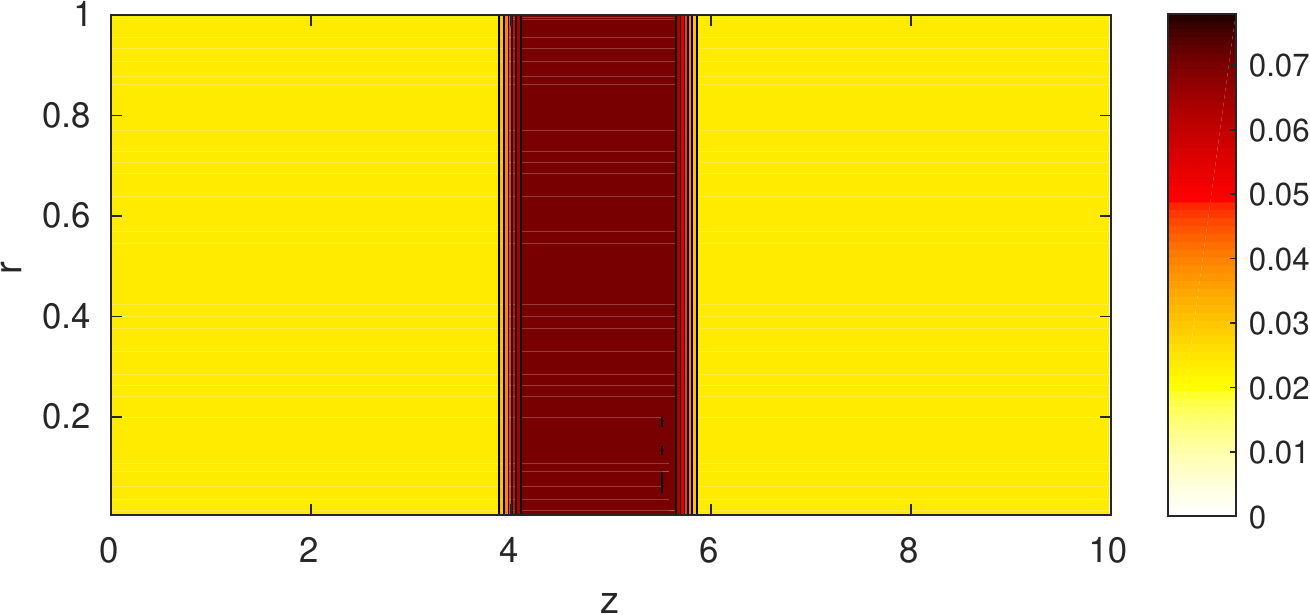} \quad 
    \includegraphics[width=0.48\textwidth]{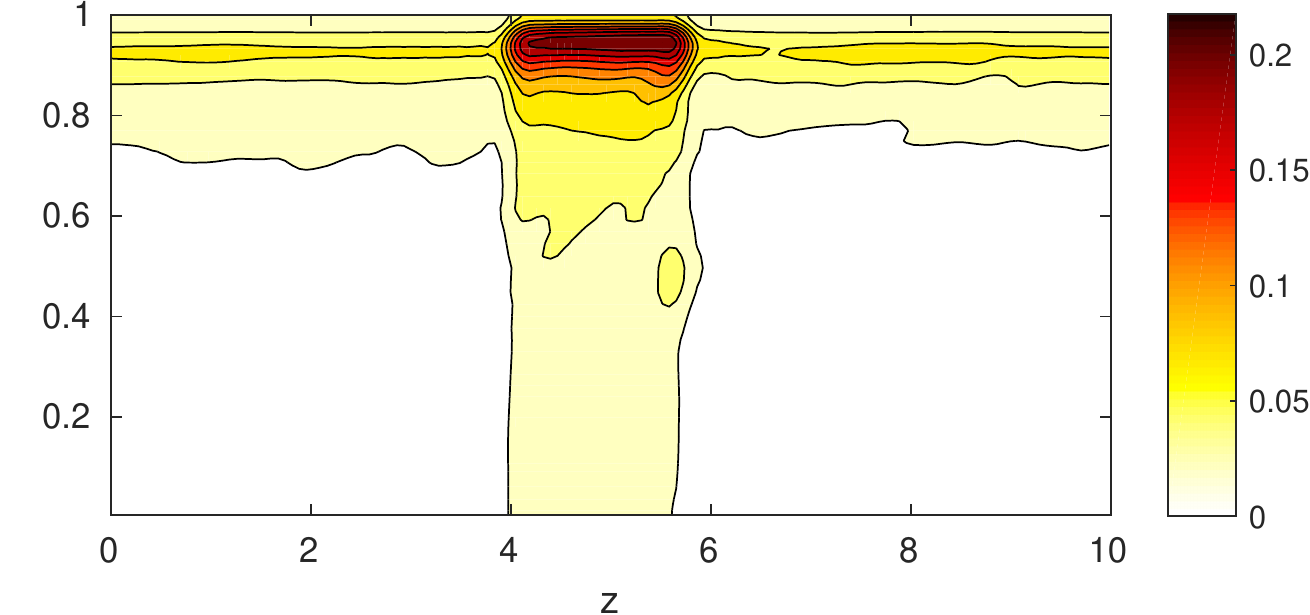}
     \caption{Cross sections in the $r-z$ plane of (left) initial guess for $\chi$ and (right) the converged $\chi$. Ten levels are used between zero and the maximum value of $\chi$. Case $L=10$, $T=300$, $\mathcal{B}_3$ as initial guess. The initial amplitude of $\chi$ is $A_0=1.1$.}
\label{fig-verythin}
\end{figure}

We start from $N=20$ turbulent fields at $Re=3000$ and, for the case with $L=10$, we consider the two \Rthree{streamwise} modulations $\mathcal{B}_1$ and $\mathcal{B}_3$,
with the amplitude appropriately rescaled to the desired $A_0$. Figures \ref{fig-verywide} and \ref{fig-verythin} show the cross sections in the $r-z$ plane of the initial guesses for $\chi$ (left) and of the converged structure of $\chi$ (right) at the end of the optimisation cycle.
In both cases, we observe that $\chi$, which initially is $r-$independent
develops a marked
radial dependence and is
concentrated close to the wall. The streamwise extent of the domain occupied by the baffle, by contrast, has changed little from the initial guesses fed into the algorithm. This may be a reflection of the fact that $\max_{\boldsymbol{x}}(-\sigma)$ is only weakly dependent, if at all, on the streamwise coordinate.


\begin{figure}
  \centering
	\includegraphics[width=1\textwidth]{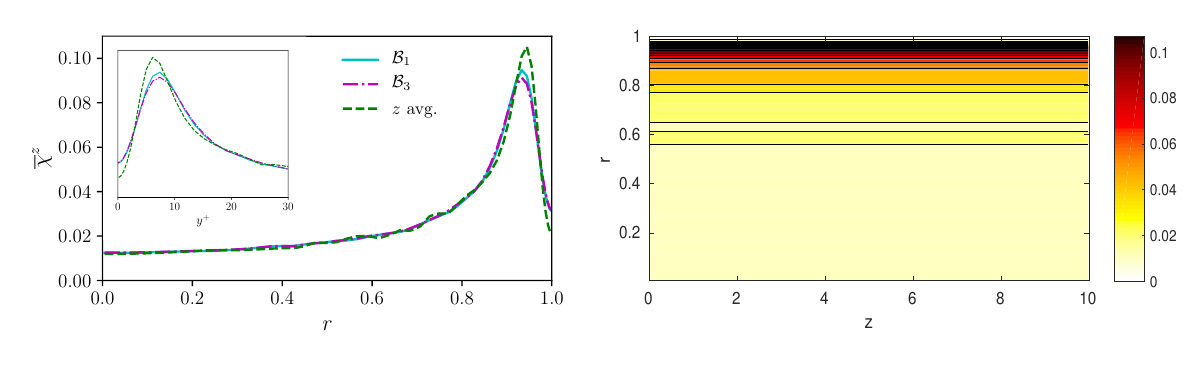}
     \caption{Case $L=10$, $T=300$, $A_0=1.1$. Left: optimal radial profiles of baffles with different streamwise modulations. \Rthree{Inset: same, as a function of distance from the wall, given in wall viscous units.} Right: cross section of the converged $\chi$ obtained from an optimisation performed with all the modes $(k,m) \neq (0,0)$ filtered out. The radial profiles obtained in the simulations shown in figures \ref{fig-verywide} and \ref{fig-verythin} were used as initial guess.}
\label{fig-cfr-opt-radprof}
\end{figure}

The optimal radial profiles $\overline{\chi}^z(r)$ for the two cases above are displayed in the left graph of figure \ref{fig-cfr-opt-radprof} and they appear to be strikingly similar. 
In both cases the peak occurs at a radial location $r \approx 0.93$,
\Rthree{corresponding to a distance from the wall, in viscous wall units, of $y^+ =Re_{\tau}(1-r)\approx 7$ (see inset), where 
$Re_{\tau} \approx 110$ in the unforced case. 
The peak is thus located in the lower part of the buffer layer ($5<y^+<30$), where the majority of the turbulent-kinetic-energy production is found to occur in DNS of wall-bounded shear flows \citep{kim-etal-1987,popebook}.
Note also that for the same Reynolds number $Re=3000$ studied here, \cite{budanur-etal-2020} found that the maximum of the turbulent-kinetic-energy production approaches $r \approx 0.9$ as the flow becomes fully turbulent.}

These optimal radial profiles found for the initial guesses $\mathcal{B}_1$ and $\mathcal{B}_3$ are then fed into our algorithm and the optimisation performed with all the modes $k>0$ filtered out, i.e. $\chi$ is restricted to a hypersurface of streamwise-independent functions of space. The $r-z$ cross section of the resulting optimal $\chi$ is shown in figure \ref{fig-cfr-opt-radprof} (right) and its radial profile is added to the left graph of figure \ref{fig-cfr-opt-radprof} for comparison. The shape of the latter profile is consistent with the other two profiles obtained with different streamwise modulations of the baffle, with the peak being even more pronounced in the streamwise-averaged case than in the previous two cases.

\begin{figure}
  \centering
	\includegraphics[width=1\textwidth]{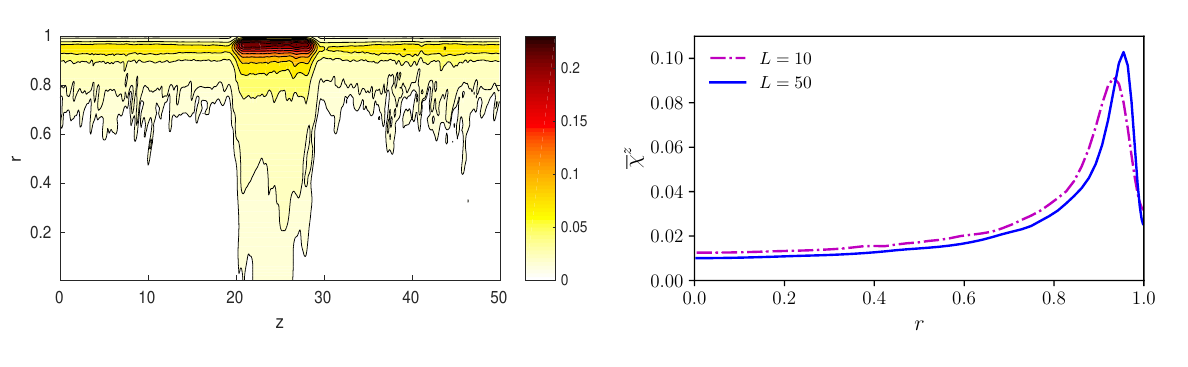}
     \caption{Cross section in the $r-z$ plane (left) and radial profile (right) of the optimal $\chi$ obtained in the case $L=50$, $T=100$, $A_0=5$. The profile obtained with $L=10$ is added to the right graph for comparison. The optimal solution is converged from a `stretched' version (i.e. $z_{start} = 20$, $z_{end} = 30 $, $\Delta z_{rise} = \Delta z_{fall} =2.5$ and $b=1/3$) of the baffle shown in figure \ref{fig-verythin}(left) which had streamwise modulation $\mathcal{B}_3$.}
\label{fig-cs-L25D-Nfield20-2}
\end{figure}

The above calculations show a tendency of the algorithm `to take material' from the middle of the pipe, move it close to the wall and then spread it more or less uniformly along the pipe. This is due to the approximate axial symmetry of $\sigma$, which is in turn due to the fast advection in a short pipe. We thus tried a longer pipe, $L=50$, which, however, was still too short \Rone{to disrupt the axial invariance of $\sigma$}, at least for optimisation problem 1. Indeed, the results shown in figure \ref{fig-cs-L25D-Nfield20-2} are analogous to those obtained with $L=10$ and $T=300$. The optimal radial profile is very similar to its $L=10$ counterpart, as shown in figure \ref{fig-cs-L25D-Nfield20-2} (right).

Ultimately, we expect the algorithm to find a streamwise localised baffle, if the pipe is sufficiently long. To ensure convergence, the optimisation needs to be started with an initial amplitude $A_0$ sufficiently large \Rone{to make the system `not too turbulent' (by ensuring all trajectories eventually decay)}.
 For this relatively large $A_0$ our calculations appear to be weakly sensitive to the $z$ support, i.e. we can have `more material' concentrated in a shorter strip of the pipe, or `less material' spread along the pipe, in both cases localised close to the wall. We might expect the algorithm to pick up the `optimal,' more localised solution, as we gradually decrease $A_0$. However, due to 
the insensitivity to the streamwise structure described above, the algorithm quickly stagnates once it has found the optimal radial profile. Only small adjustments to the radial profile are sufficient to keep the flow laminar as $A_0$ is gradually decreased. Unexpectedly, this is also the case in the $L=50$ pipe, as shown in figure \ref{fig-cs-L25D-Nfield20-2}. If $A_0$ is decreased too rapidly, a baffle that does not relaminarise the flow might be encountered, thus preventing convergence. 


Solving the optimisation problem as a decreasing function of $A_0$
is time-consuming because $A_0$ needs to be decreased in small steps in order to ensure convergence.
This procedure was carried out for the streamwise-averaged case in the $L=10$ pipe. Starting from $A_0=1.1$ (see figure \ref{fig-cfr-opt-radprof}) we were able to decrease the initial amplitude down to $A_0=0.7$. The optimal radial profile obtained in the latter case is shown in figure \ref{fig-fitting2} and its shape is found to have changed very little from the case with $A_0=1.1$ (compare green dashed lines in figures \ref{fig-cfr-opt-radprof}(left) and \ref{fig-fitting2}(left)). Therefore, we did not carry out the same procedure for the $L=50$ case which is more computationally expensive. 

\subsection{`Manual' localisation}
\label{sec:res_manual}
\begin{figure}
  \centering
      \includegraphics[width=0.48\textwidth]{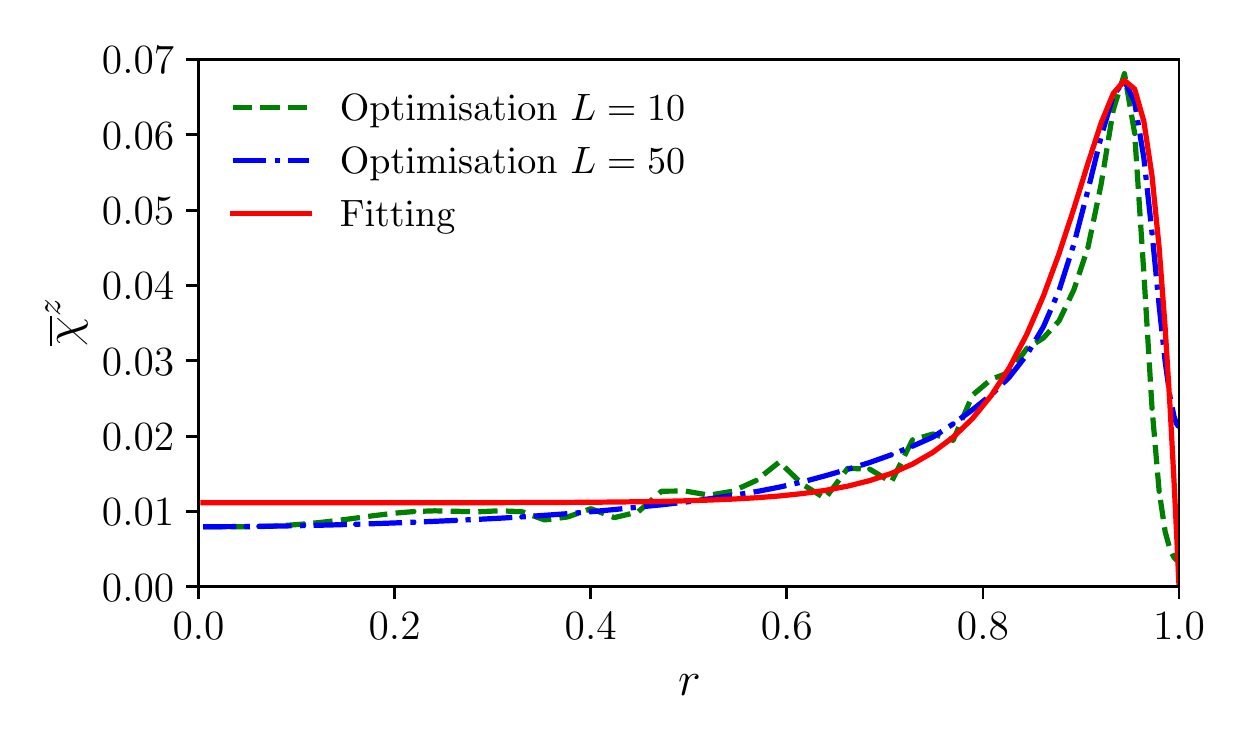} 
      \includegraphics[width=0.48\textwidth]{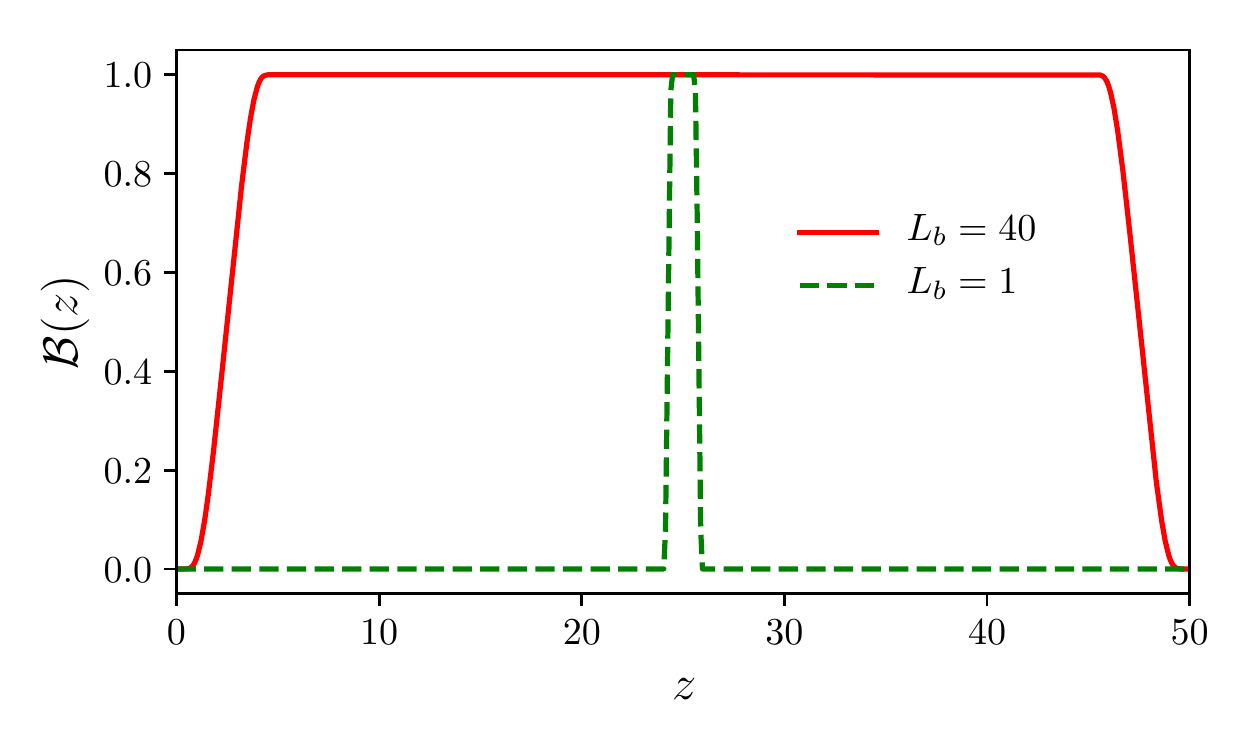} 
     \caption{Left: Analytical fitting of the optimal radial profiles, obtained from a streamwise-averaged optimisation in both $L=10$ and $L=50$ cases. For the $L=10$ case we started from the optimal $\chi$ obtained with $A_0=1.1$ (see figure \ref{fig-cfr-opt-radprof}) and gradually decreased the amplitude to $A_0=0.7$. The same procedure was not carried out for $L=50$ due to the computational expense. Since the radial profile only undergoes small adjustment when $A_0$ is gradually decreased, a rescaled version from $A_0 \approx 5$ is used in the $L=50$ case. Right: Streamwise modulation of the baffle for the cases $L_b=40$ ($z_{start} = 0$, $z_{end} = 50 $, $\Delta z_{rise} = \Delta z_{fall} =5$ and $b=0$) and $L_b=1$ ($z_{start} = 24$, $z_{end} = 26 $, $\Delta z_{rise} = \Delta z_{fall} =0.5$ and $b=0$). Note that $b>0$ was only needed to initialise the optimisation in \S \ref{subsec:res_IG} and \S \ref{sec:result_pb1}, while here we use $b=0$ to study the effect of the streamwise localisation of the baffle.}
\label{fig-fitting2}
\end{figure}

To better understand the optimal streamwise structure of $\chi$, we use an analytical fitting $f(r)$ for the optimal radial profile found in \S \ref{sec:result_pb1} and perform a parametric study on the effect of the streamwise extent $L_b$ of the baffle. 
\Rthree{The procedure amounts to writing $\chi(r,z)$ as the product $f(r)\, \mathcal{B}(z)$, up to a suitable rescaling factor for adjusting the amplitude to the desired one, where $f$ is fixed by \eqref{eq-fitting} and $\mathcal{B}(z)$ is optimised `manually'.}
We consider the optimal radial profiles obtained in the $L=10$ and $L=50$ pipes for the streamwise-averaged cases
  and fit the following curve
\beq
f(r) = [c_1(e^{c_2 r}-1)+c_3]\times[\tanh((1-r)/c_4)]
\label{eq-fitting}
\eeq
\Rthree{where $c_i$, $i=1$ to $4$, are free parameters.
Some experimentation with $c_i$ 
 showed that the choice $c_1=0.0000015$, $c_2=12.75$, $c_3=0.035$ and $c_4=0.06$
was the most robust at suppressing turbulence, that is, the flow could be kept laminar with lower values of $A_0$, thus improving the performances of the baffle. The resulting $f(r)$ is shown in figure \ref{fig-fitting2}(left).   
In particular, we noticed that, while the optimal radial profile of $\chi$ is concentrated close to the wall, a non-zero $\chi$ is needed close to the centre of the pipe in order to relaminarise the flow.
The \Rthree{amplitude of the flat part} of $\overline{\chi}^{z}(r)$ for $0 \leq r\lesssim 0.5$ was found to be crucial for the suppression of turbulence, i.e. a small decrease in it would cause the baffle to fail at relaminarising the flow. 
Reasons why the algorithm did not find the analytical fitting shown in figure \ref{fig-fitting2} as the optimal may be ascribed to the convergences issues encountered while gradually decreasing $A_0$, as discussed in \S \ref{sec:result_pb1}, and to the ensemble averaged being limited to $N=20$ because of the computation cost.}
\begin{figure}
  \centering
      \includegraphics[width=0.48\textwidth]{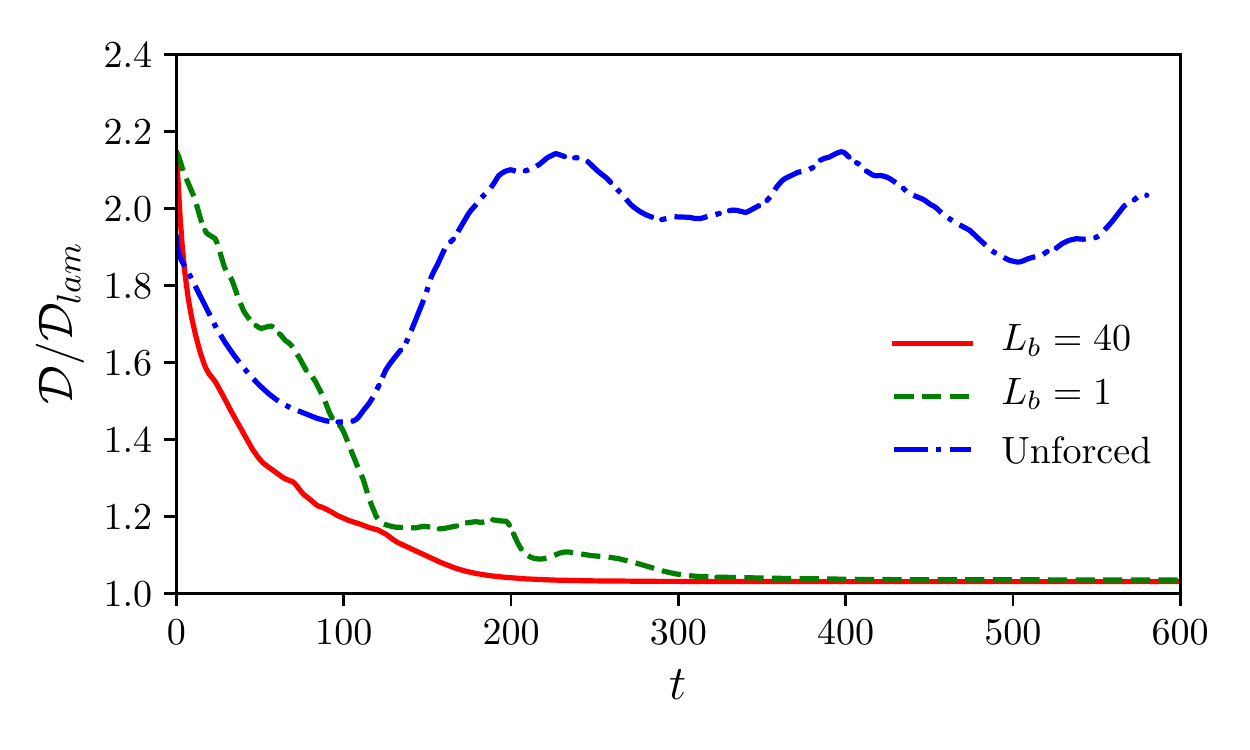} 
      \includegraphics[width=0.48\textwidth]{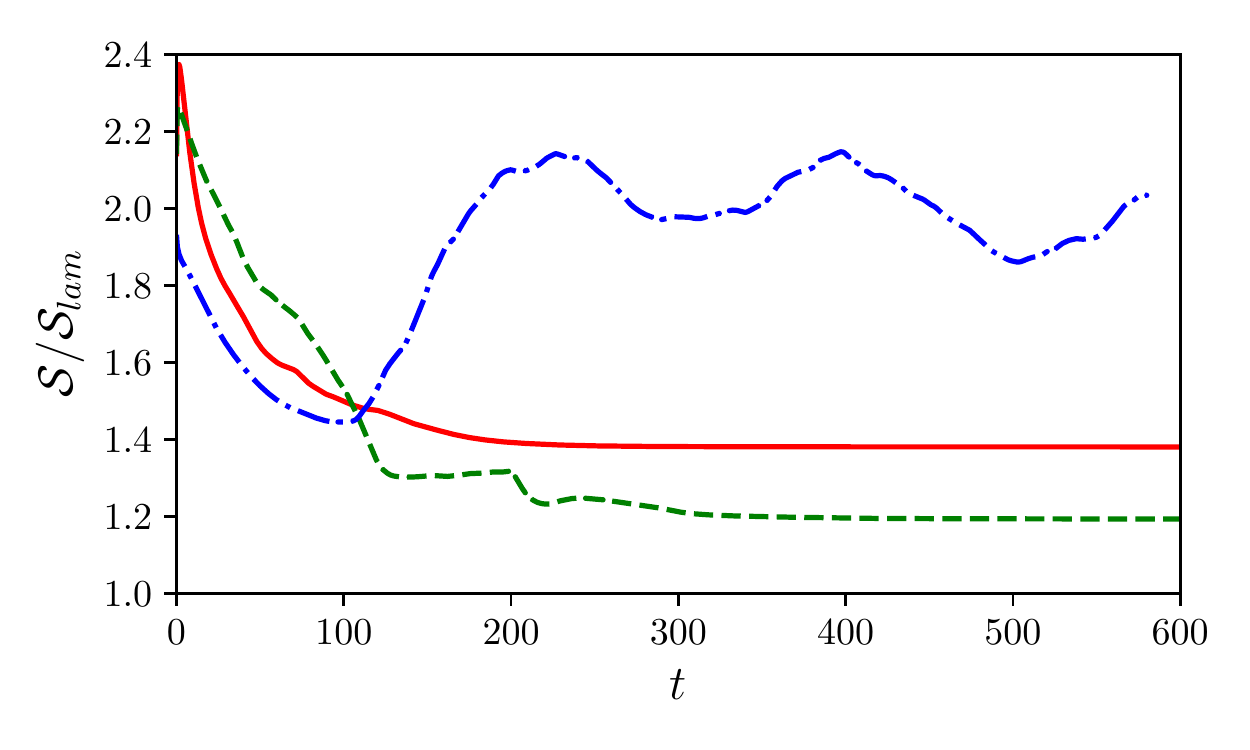}
     \caption{Effect of the baffle length $L_b$ for a $50R$-long pipe. Time series of (left) dissipation and (right) shear stress at the wall for different streamwise modulations of the baffle at the critical amplitude $A_0=A_{crit}$ that can just relaminarise the flow.}
\label{fig-manualzoptimis-longpipe}
\end{figure}
\begin{figure}
  \centering
    \includegraphics[width=0.6\textwidth]{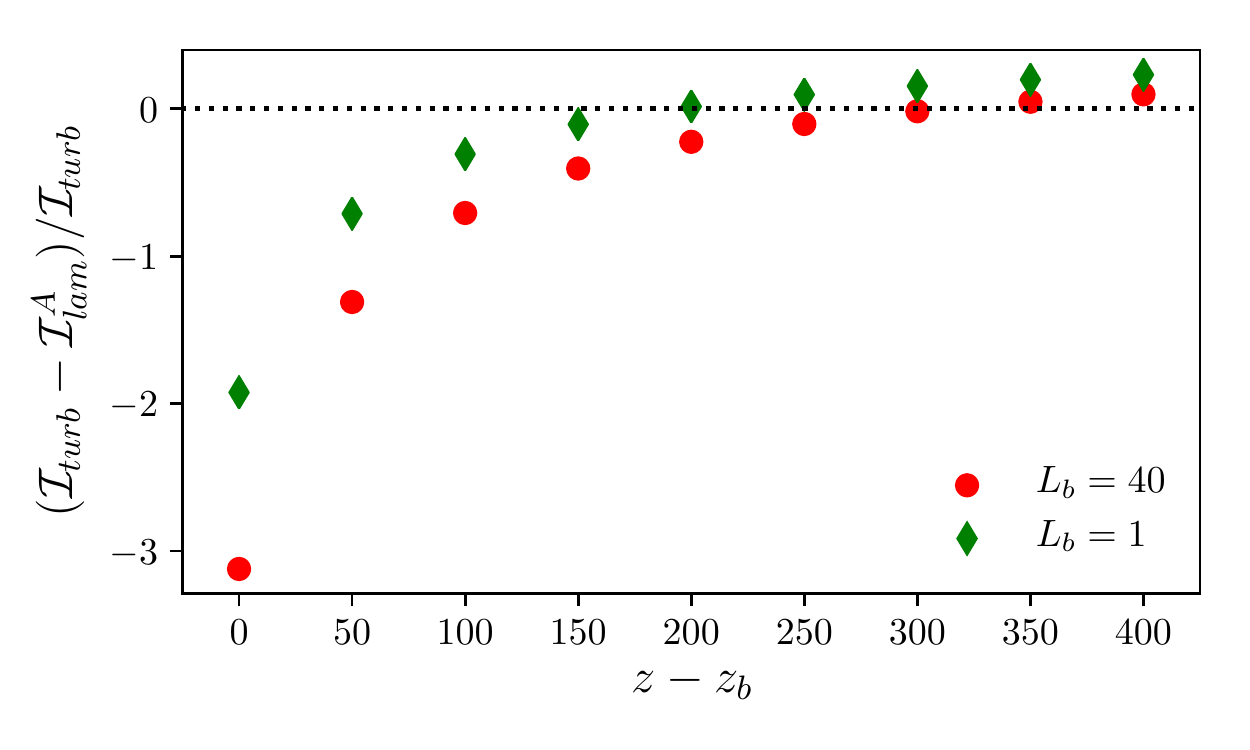}
     \caption{Net energy saving due to the optimised baffle at $Re=3000$, $L=50$ for two lengths $L_b$ of the baffle as a function of the streamwise distance $z$ from the baffle location $z_b$. The net energy is measured as a relative difference between the input energy needed to drive the flow in the turbulent unforced case, $\mathcal{I}_{turb}$, and in the relaminarised forced case $\mathcal{I}_{lam}^A$. The crossing point of each curve with the zero corresponds to the break-even length $L_{even}$.}
\label{fig-OptF-betacorr-Re3000}
\end{figure}

The results are only shown for the long-pipe case, for which the effect of the baffle localisation is more evident and, for the rest of the paper, unless otherwise specified, we will assume $L=50$. Several streamwise extents and modulations were tested, of which here we present results only for the two, most representative, cases: $L_b=40$ and $L_b=1$, shown in figure \ref{fig-fitting2} (right).
 Figure \ref{fig-manualzoptimis-longpipe} displays the time series of dissipation and wall shear stress for $L_b=40$ and $1$, as well as for the unforced case. The simulations were fed with $N=5$ different turbulent initial conditions, although the results are presented here only for one of them. For both streamwise extents of the baffle the initial amplitude was decreased until relaminarisation was not possible any more for at least one turbulent field. We found that the critical amplitude $A_{crit}$ that can relaminarise all five turbulent initial conditions is $A_{crit} = 4.3$, for $L_b=40$, and $A_{crit} = 2.9$, for $L_b=1$. Comparison of the long-time asymptotes for the two cases displayed in figure \ref{fig-manualzoptimis-longpipe} shows that the dissipations decay to an almost identical value,  while the 
wall shear stress
 reached with the short baffle after relaminarisation is about half (after subtracting off the laminar unforced value of 1) that obtained with the long baffle. The shear stress reductions at the wall, indeed, are approximately $20\%$ and $40\%$ with the long and short baffle, respectively. This results in a lower input energy needed with the short baffle, compared to the long one. In both cases, the energy spent to compensate for the local pressure drop $\mathscr{B}$ (defined in \eqref{eq-NSz}) immediately downstream of the baffle is larger than the energy saved by relaminarising the flow. However, assuming the flow remains laminar downstream of the baffle, there will be a break-even point 
 where the shear stress reduction at the wall due to the relaminarised flow compensates for the extra drag produced by the baffle. We define such break-even length $L_{even}$ as the downstream distance from the baffle where the energy inputs needed to drive the flow in the relaminarised forced case $\mathcal{I}_{lam}^A$ and in the turbulent unforced case, $\mathcal{I}_{turb}$ are equal. For a downstream distance $z>L_{even}$ a net power saving is achieved. Figure \ref{fig-OptF-betacorr-Re3000} shows that $L_{even} \approx 200$ for the short baffle and $L_{even} \approx 300$ for the long baffle. The break-even length for the short baffle is consistent with the experiments of \cite{kuhnen-etal-2018a}.

\begin{figure}
  \centering
    \includegraphics[width=1\textwidth]{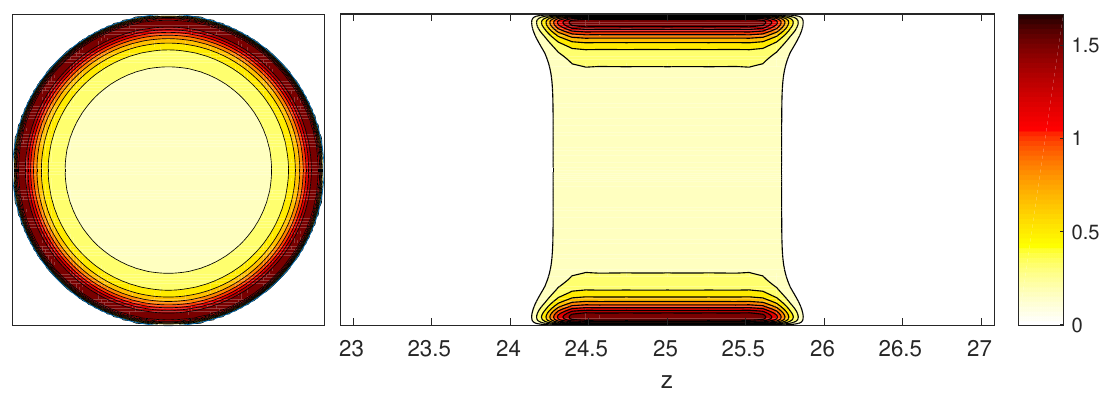}
     \caption{Optimal $\chi(r,z)$ for a $L=50$ pipe. Cross sections in the $r-\theta$ plane at $z=25$ (left), where the baffle is localised, and in the $r-z$ plane (right). Note that the baffle occupies a very small region of the pipe length and is also radially concentrated close to the pipe walls. The critical amplitude for relaminarisation is $A_{crit}=2.9$.}
\label{optF-longpipe}
\end{figure}

 For a baffle extent $L_b<1$, relaminarisation was not found to be possible. Therefore, the optimal baffle is streamwise localised and in a $L=50$ pipe the minimum extent of the baffle below which the flow cannot be relaminarised is $L_{b, min}=1$.
The optimal $\chi(r,z)$ is shown in figure \ref{optF-longpipe}.

The fact that the dissipations after relaminarisation are almost identical for the wide and localised baffles while the input energy is lower for the latter, suggests that we should minimise the input energy (optimisation problem 2) rather than the dissipation (optimisation problem 1). At least in the long-pipe case, we should expect the algorithm to converge to a streamwise localised baffle. This is discussed in the following section.
\subsection{Optimisation problem 2}
\label{sec:result_pb2}

\Rone{In optimisation problem 2 the objective functional to minimise is the total energy input.}
In the previous sections we showed that minimising the total viscous dissipation \Rone{(optimisation problem 1)} allows us to quickly find the optimal radial shape of the baffle, while clear convergence to an optimal streamwise structure was not achieved. A parametric study on the effect of the baffle extent showed that a localised baffle can reduce the input energy to a lower value than a wide baffle, the dissipations reached in both cases after relaminarisation being instead very similar. This suggests that minimising the input energy may help the algorithm to capture the streamwise localisation of the baffle and motivated us to move onto optimisation problem 2. 
However, the results obtained solving optimisation problem 2 still show 
weak dependence on the $z$-support, so we do not see streamwise localisation of $\chi$. 

\begin{figure}
  \centering
      \includegraphics[width=0.48\textwidth]{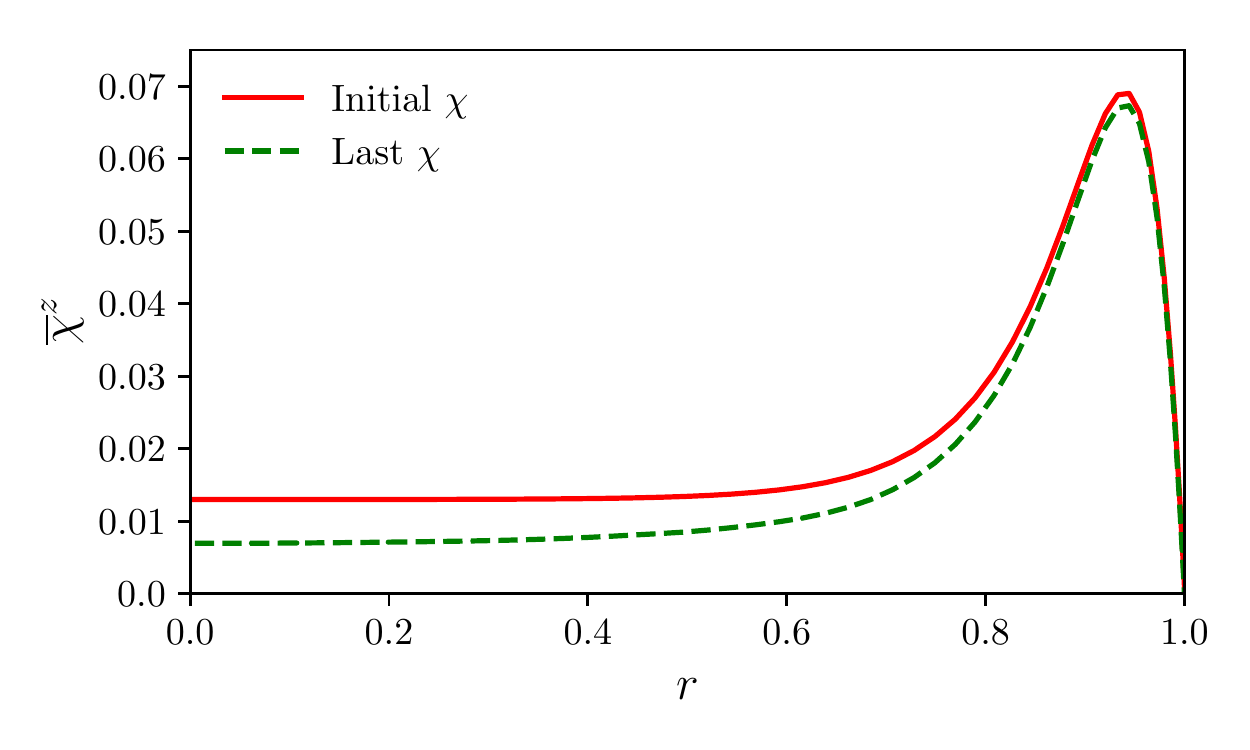} 
      \includegraphics[width=0.48\textwidth]{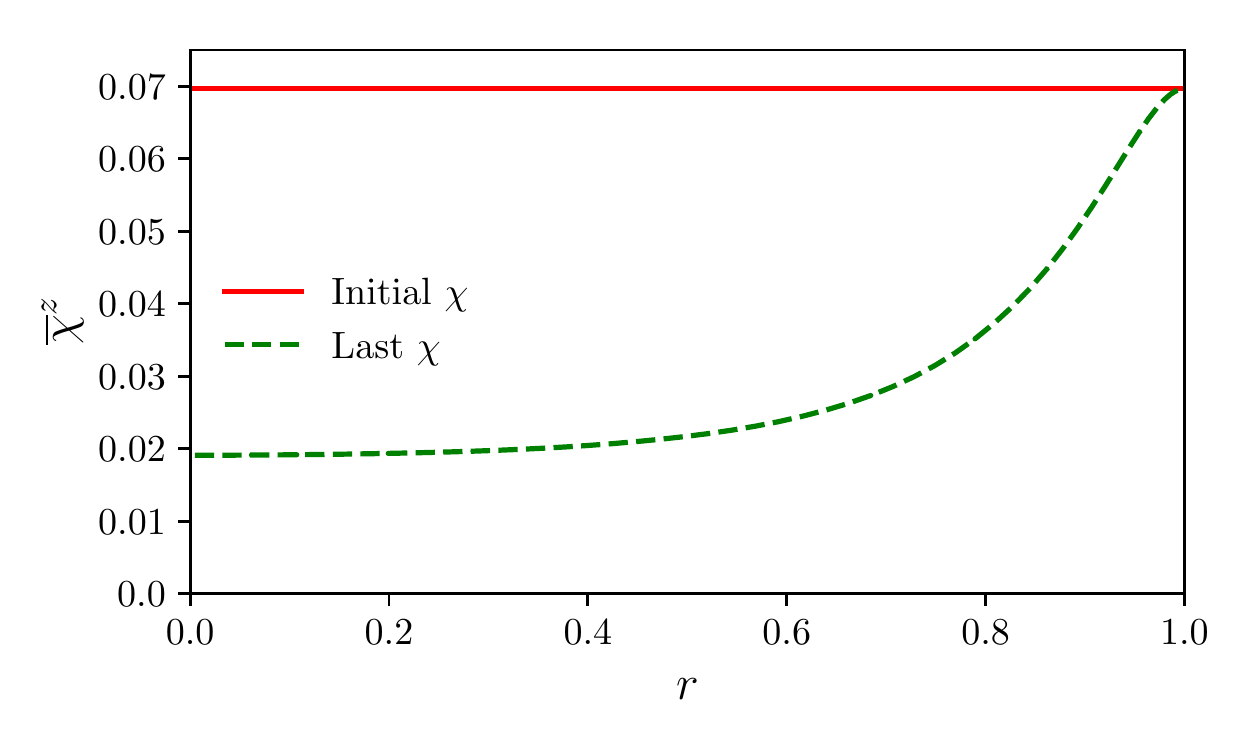}
     \caption{Radial profiles of $\overline{\chi}^{z}$ obtained by solving optimisation problem 2 (minimise the input energy) at $Re=3000$, $L=10$, starting from different initial guesses for $\overline{\chi}^{z}$. In both cases $\mathcal{B}=\mathcal{B}_3$. The last $\overline{\chi}^{z}$ is that obtained just before convergence fails, due to the issues explained in the text. Left: the initial $\overline{\chi}^{z}(r)$ was obtained from optimisation problem 1 (minimise the viscous dissipation) and it remained almost unchanged. Right: the initial $\overline{\chi}^{z}(r)$ was radially homogeneous and convergence failed before the optimal radial profile characterised by the minimal amplitude, shown in the left graph, could be reached.}
\label{fig-minI}
\end{figure}

As anticipated in \S \ref{subsec:pb2},
convergence is problematic with optimisation problem 2 because $A_0$ is allowed to vary and typically the algorithm tries to decrease it in order to minimise the work done, and thus the input energy. If $A_0$ is decreased too much in one step, then some of the turbulent initial fields can become turbulent again, thus preventing convergence due to the sensitivity to initial conditions.
When optimisation problem 2 is fed with a radially homogeneous initial guess, 
a tendency of the algorithm to `remove' material from the centre of the pipe is observed, as shown in figure \ref{fig-minI}(right). This suggests that a radial profile concentrated close to the wall, as the one obtained with optimisation problem 1 (and reported in the left graph of figure \ref{fig-minI}), would eventually be found. However, convergence failed before the optimal radial profile, characterised by the minimal amplitude, could be reached. 
Because the streamwise component of the total velocity field is larger in the core of the pipe than towards the wall, 
the work done can be decreased more efficiently by decreasing $\chi$ in the pipe centre rather than close to the wall. Thus, the trend shown in figure \ref{fig-minI} is not surprising. 
 For the same reason, before convergence could be reached, it is also not too surprising that $\chi$ remains essentially unvaried at the wall (where $\uvec_{tot}=\mathbf{0}$), as shown in figure \ref{fig-minI}(right).
 
On the other hand, if fed with a good initial guess for $\chi$ (for example obtained from optimisation problem 1), optimisation problem 2 is able to provide $A_{crit}$ (or very close to it) in a few iterations, much more quickly than by gradually decreasing $A_0$ as done in optimisation problem 1.
The shape of the optimal radial profile obtained from optimisation problem 1 remains almost unchanged when fed into optimisation problem 2, as shown in figure \ref{fig-minI}(left), although, for the reason noted above, we observe a tendency of the algorithm to slightly decrease $\chi$ in the region close to the centre of the pipe.
As noticed earlier, the convergence issue encountered with optimisation problem 2 is ascribed to the
 tendency of the algorithm to reduce $A_0$ globally
 in order to reduce the work done. Furthermore, in \S \ref{sec:res_manual}, while experimenting with different radial fittings for the optimal $\chi$, we noticed that the relaminarisation process is very sensitive to the value of $\chi$ near the centre, where the algorithm tends to decrease $\chi$ the most.
This behaviour adds to the convergence difficulties presented by optimisation problem 2.

In summary, when fed with a good initial guess for $\chi$, optimisation problem 2 is able to decrease the baffle amplitude 
 to values very close to those obtained `manually' (see \S \ref{sec:res_manual}) for the same extent of the baffle. Therefore, this study, although not able to find the optimal streamwise-localised shape,
 helped us confirm the values of $A_{crit}$ obtained `manually' in \S \ref{sec:res_manual} for the different baffle extents.

\section{Mean profiles}
\label{sec:meanprof}
\begin{figure}
  \centering
    \includegraphics[width=0.6\textwidth]{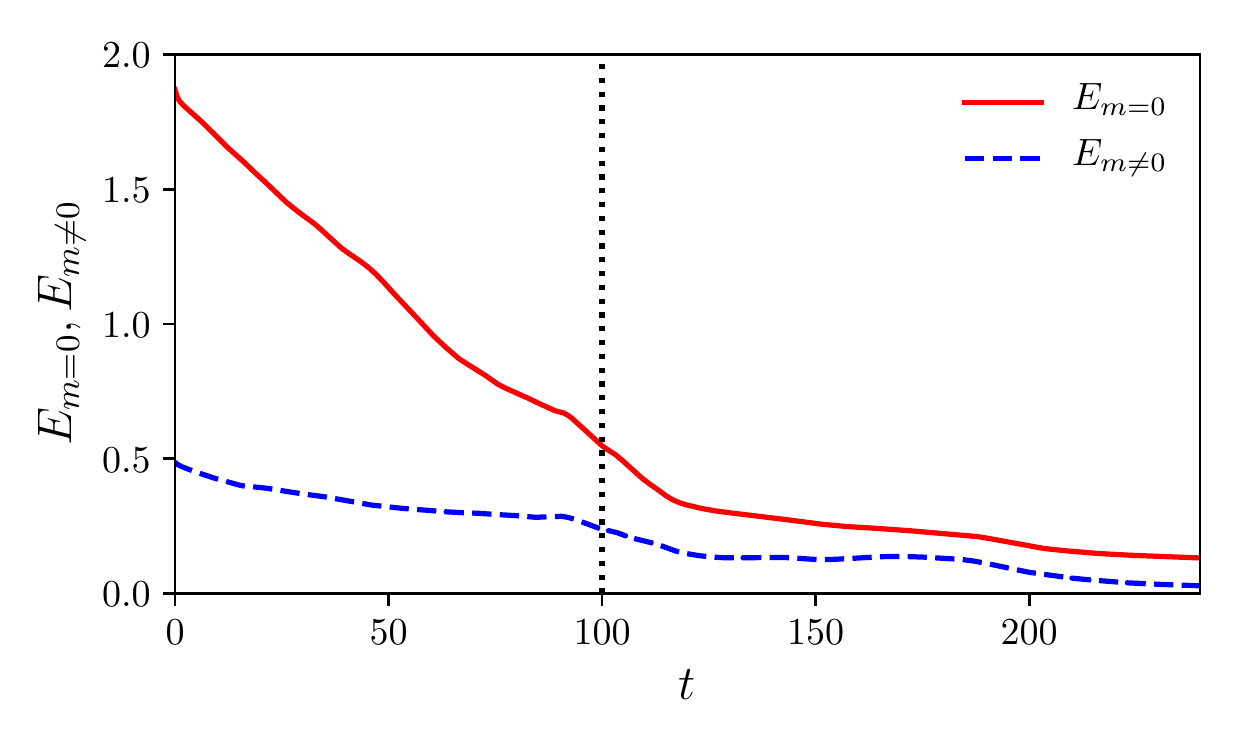}
     \caption{ $Re=3000$, $L=50$. Time series of energy in the $m=0$ and $m\neq 0$ modes using the optimal baffle shown in figure \ref{optF-longpipe}. Almost all of the relaminarisation happens in the first $t=100-120$ time units. The vertical line indicates the time horizon $T=100$ used in the optimisation.}
\label{optF-Em}
\end{figure}
\begin{figure}
\hspace{-1.7cm}
    \includegraphics[width=1.3\textwidth]{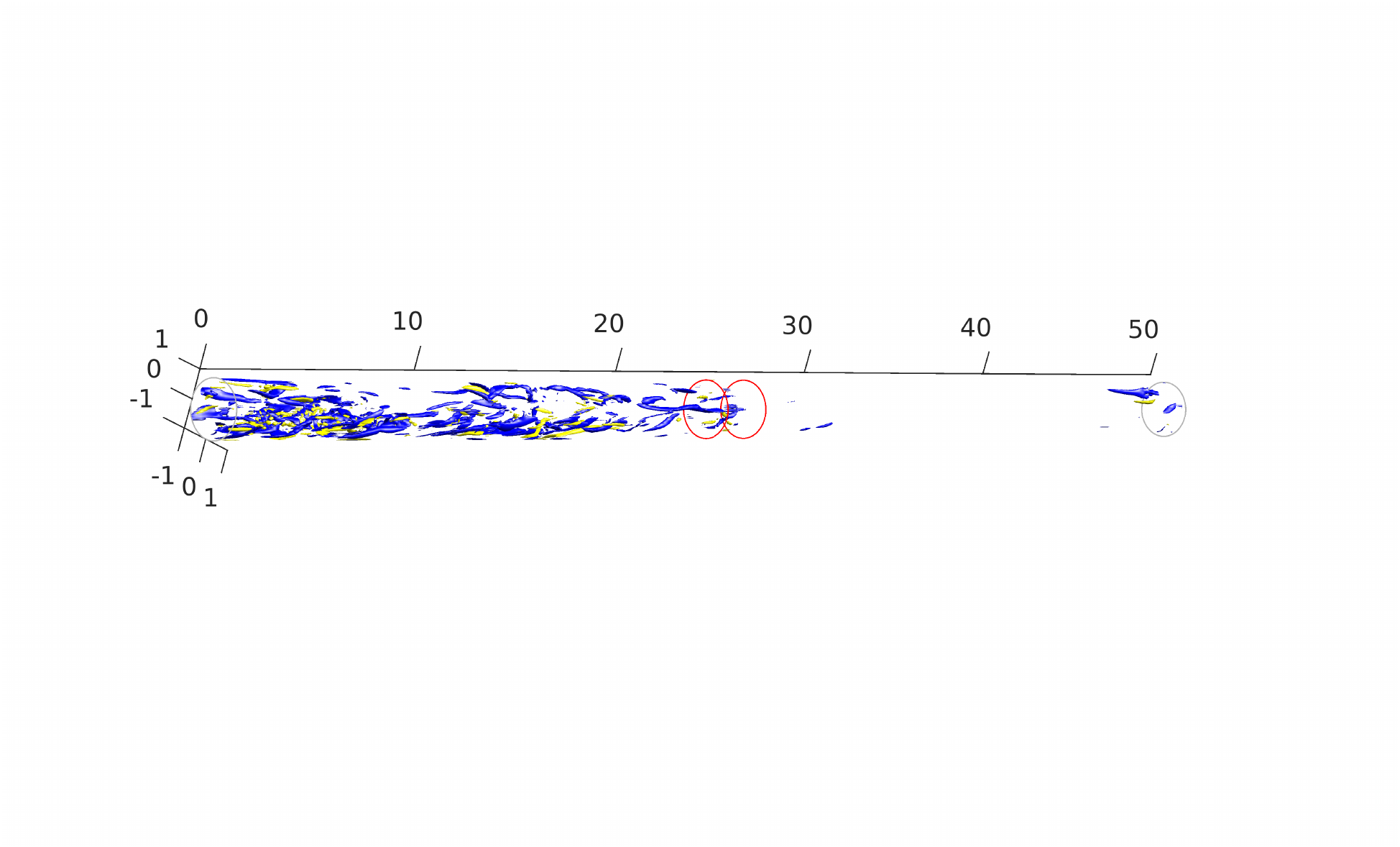} 
\vspace{-4cm}
     \caption{$Re=3000$, $L=50$. Iso-contours of streamwise vorticity (light/yellow and dark/blue indicate $20\%$ of the maximum and minimum, respectively) at $t^{\dagger}= 60$ using the optimal baffle shown in figure \ref{optF-longpipe}. At this time, approximately half of the fluid has passed through the baffle. The region occupied by the baffle is indicated with red circles.}
\label{optF-isocont}
\end{figure}
\begin{figure}
	\includegraphics[width=1\textwidth]{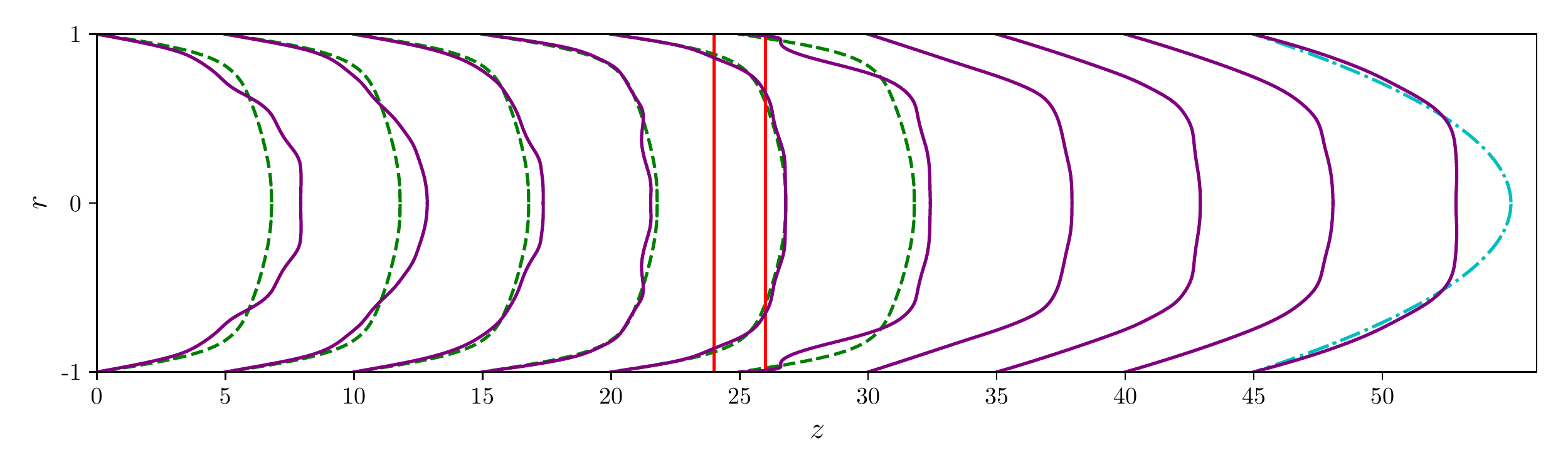} 
     \caption{ $Re=3000$, $L=50$. Mean velocity profiles along the pipe at $t^{\dagger}= 60$ using the optimal baffle shown in figure \ref{optF-longpipe}. At this time, approximately half of the pipe flow has relaminarised. The region occupied by the baffle is indicated with red vertical lines. At $z=25$ (in the middle of the baffle) we can see little kinks close to the walls. The mean profile at $z=45$ is approaching the parabolic profile (indicated by the cyan dash-dotted line). The green dashed profiles represents the reference (turbulent unforced) mean profile $U_{ref}:=\overline{u_{z,tot}}^{\theta,z}(t=0)$.}
\label{optF-meanprof}
\end{figure}

In order to link our results to the experiments of \citet{kuhnen-etal-2018a}, we fix $\chi(r,z)$ to be the optimal structure shown in figure \ref{optF-longpipe} and analyse how such a baffle modifies the mean streamwise velocity profiles.
Figure \ref{optF-Em} shows the time series of the energy contained in the azimuthal-independent and azimuthal-dependent modes in the long-pipe case $L =50$. Almost all of the relaminarisation happens in the first $t=100$ time units, that is, in the first pass through the baffle. Indeed, approximately halfway through this time ($t \approx 50-60$), almost exactly half of the pipe is turbulent and the other half laminar, as shown in the iso-contours of figure \ref{optF-isocont}.
At $t^{\dagger} = 60$ we analyse the mean streamwise velocity profiles along the pipe (see figure \ref{optF-meanprof}) and compare them with the experimental results of \citet{kuhnen-etal-2018a} (refer to their figure 7).
Note that the baffle is at $24.5 - 25.5$, as shown in figure \ref{optF-longpipe} and marked with vertical lines in figure \ref{optF-meanprof}.
 The incoming flow at $z=20$ is very similar to the reference (unforced) mean turbulent profile $U_{ref}:=\overline{u_{z,tot}}^{\theta,z}(t=0)$. The mean profile at $z=25$ (in the middle of the baffle) features little kinks close to the walls and, further downstream, the profile at $z=45$ is approaching the parabolic shape. In comparison to figure 7 of \citet{kuhnen-etal-2018a}, our profiles do not present the overshoots close to the wall, which cause a big pressure drop just downstream of the baffle. Our optimal $\fvec(\mathbf{x},t)$ indeed avoids these M-shaped profiles by being concentrated close to the wall.

\begin{figure}
  \centering
    \includegraphics[width=0.48\textwidth]{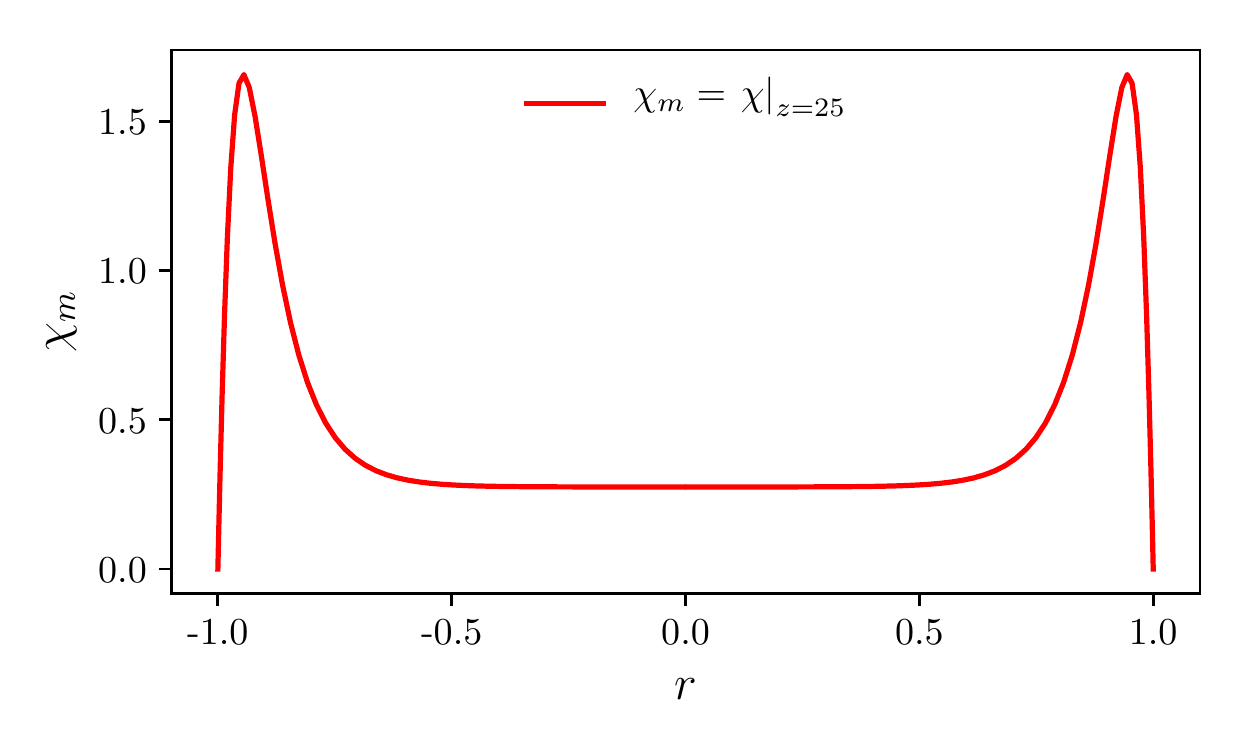}\\
    \includegraphics[width=0.48\textwidth]{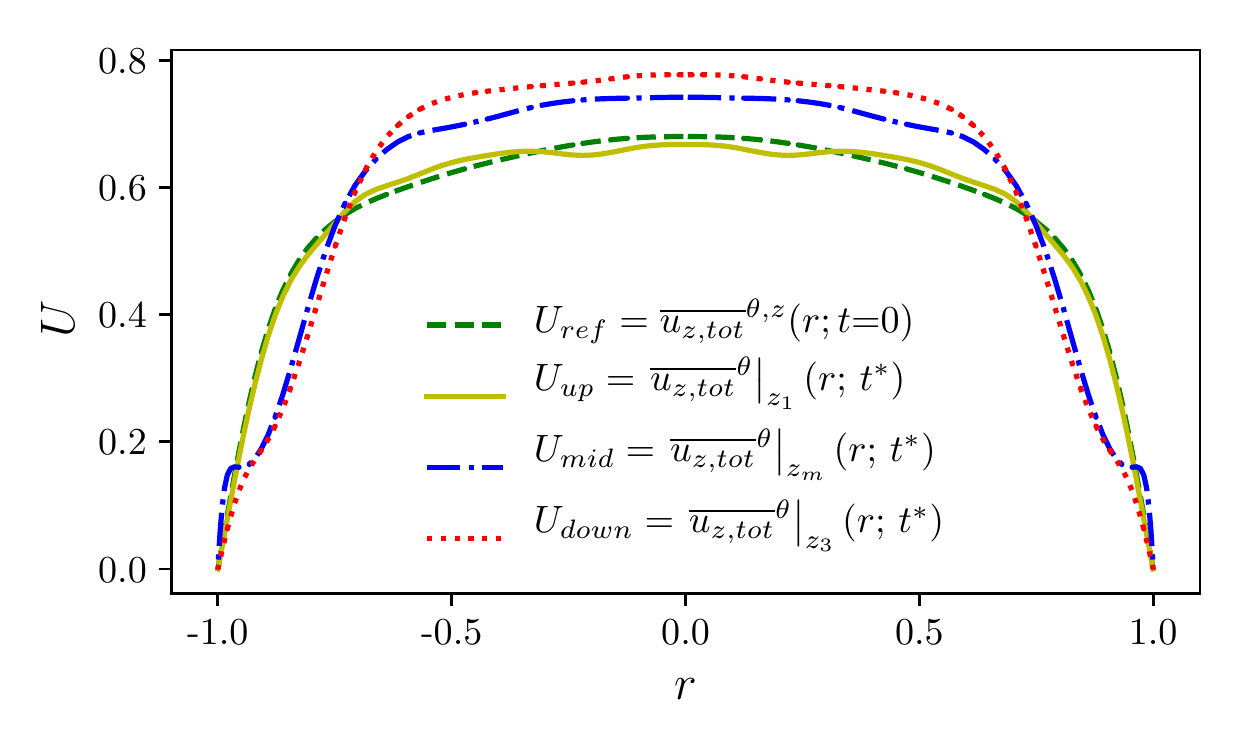}
    \includegraphics[width=0.48\textwidth]{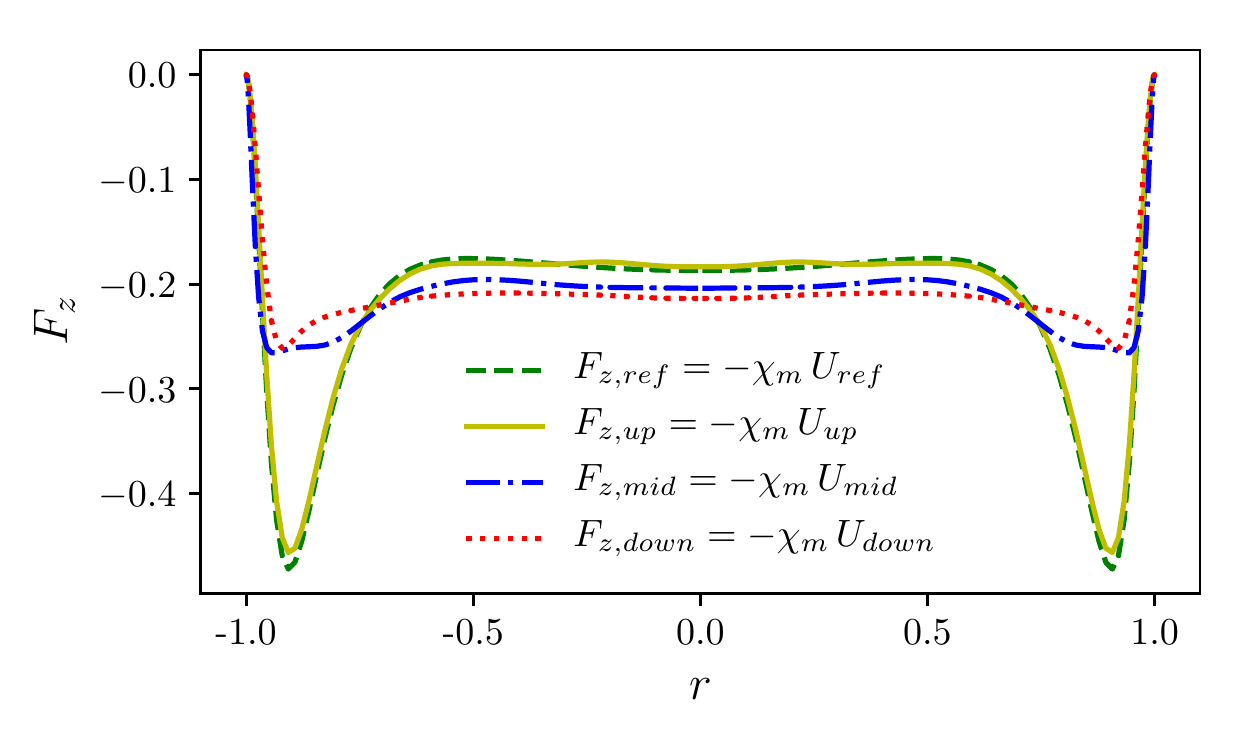}
     \caption{$Re=3000$, $L=50$, the baffle is at $24.5 - 25.5$ (locations indicated with red vertical lines in figure \ref{optF-meanprof}). The spatial distribution of $\chi(r,z)$ is fixed to be the optimal shown in figure \ref{optF-longpipe} and its radial profile $\chi_m(r):=\left.\chi\right|_{z=z_m}$ in the middle of the baffle $z_m=25$ is shown in the top graph. Bottom: (left) streamwise velocity profiles $U(r)$ and (right) streamwise component of the drag force $F_z(r) = -\chi_m\, U(r)$ at different streamwise locations and times.  
At $t^{\dagger}= 60$ approximately half of the fluid has passed through the baffle and half of the pipe has relaminarised. The radial distributions of drag just upstream/downstream ($z_1=24$, $z_2=26$) of the baffle and in the middle are: $F_{z,up}=-\chi_m\,U_{up}$, $F_{z,down}=-\chi_m\, U_{down}$ and $F_{z,mid}=-\chi_m\, U_{mid}$, respectively, where $U_{up}=\left.\overline{u_{z,tot}}^{\theta}\right|_{z=z_1}$, $U_{down}=\left.\overline{u_{z,tot}}^{\theta}\right|_{z=z_2}$, $U_{mid}=\left.\overline{u_{z,tot}}^{\theta}\right|_{z=z_m}$ at $t=t^*$.
At $t=0$ the effect of the baffle has not been felt yet and the incoming flow is fully turbulent.
 The baffle drag can be expressed as $F_{z,ref}=-\chi_m\,U_{ref}$, where $U_{ref}(r):=\overline{u_{z,tot}}^{\theta,z}(t=0)$ is the reference (unforced) mean turbulent profile.}
\label{optF-forcing-prof}
\end{figure}

To better understand the effect of the baffle drag $\fvec(\mathbf{x},t)= -\chi(r,z)\,\mathbf{u}_{tot}(\mathbf{x},t)$ on the flow, we analyse the radial profiles of its streamwise component $F_z=\fvec \cdot \hat{\mathbf{z}}$, see figure \ref{optF-forcing-prof}.
The radial profile of $\chi$ is fixed to be the optimal shown in figure \ref{optF-longpipe} at the midpoint of the computational domain, i.e. $\chi_m(r):=\left.\chi\right|_{z=25}$ (see top graph of figure \ref{optF-forcing-prof}). 
As noted above, the turbulent incoming flow upstream of the baffle $U_{up}(r)=\left.\overline{u_{z,tot}}^{\theta}\right|_{z=24}(r;\,t^*)$ is similar to the reference mean turbulent profile $U_{ref}(r)$ (see bottom left graph). In both cases, the resulting baffle drags $F_{z,ref}(r)=-\chi_m\,U_{ref}$ 
and $F_{z,up}(r)=-\chi_m \, U_{up}$ (see bottom right graph) are characterised by a very flat profile in the middle and present overshoots close to the wall, 
possibly in order to kill the near-wall turbulence regeneration cycle. 
It should be noted that a full collapse of turbulence in a pipe flow was also obtained in the numerical experiments of \citet{kuhnen-etal-2018b} by introducing a body force $\tilde{\fvec}=\tilde{F}(r)\hat{\mathbf{z}}$ that flattens the mean streamwise velocity profile. 
It is worth remarking, however, that such body force was negative (pointing upstream) in the centre and positive (pointing downstream) near the wall, i.e. it decelerates the flow in the core and accelerates it close to the wall, with the mass flux kept constant (refer to their figure 7 in the Extended Data). In contrast, our baffle drag is always positive everywhere, i.e. it only removes energy from the flow, as shown in figure \ref{optF-forcing-prof}, so it is realisable with a passive control method. Furthermore, our baffle drag is strongly localised in the streamwise direction while \citet{kuhnen-etal-2018b}'s forcing was imposed globally on top of a fully turbulent flow and is thus impractical to implement in experiments. \Rthree{For this reason, \cite{song-2014} (see their section 4.1.6) also considered a `chopped' version of such forcing and showed that the streamwise localised forcing is still capable to suppress turbulence}. 

\section{Dependence on the Reynolds number}
\label{sec:result_Re}
The solution of optimisation problem 1, combined with a parametric study on the baffle extent $L_b$, provided the optimal baffle design at $Re=3000$.
 This optimal baffle was found to be axisymmetric, radially concentrated close to the wall and streamwise localised, with a minimal length of one radius and a critical amplitude for relaminarisation $A_{crit}=2.9$ in a $L=50$ pipe (refer to figure \ref{optF-longpipe}).
We now investigate, \EM{via DNS}, whether such baffle, after appropriate rescaling of the amplitude, can relaminarise the flow at higher Reynolds numbers. 
 We consider four Reynolds numbers $Re=5000$, $7000$, $10\,000$ and $15\,000$ and $N=5$ turbulent initial conditions for each of them (except at $Re=15\,000$ for which only one turbulent initial field was considered to limit the computational cost) \EM{and $L=50$ throughout}.

In the following, we define a skin-friction drag reduction in the laminar and turbulent cases as:

\begin{equation}
\mathcal{FR}_{lam/turb}:=\frac{(1+\mathscr{T}_w)_{turb}-(1+\mathscr{T}_w)^{A}_{lam/turb}}{(1+\mathscr{T}_w)_{turb}}=\frac{\mathcal{S}_{turb} - \mathcal{S}_{lam/turb}^{A}}{\mathcal{S}_{turb}}\,,
\label{DRsk}
\end{equation}
where $(1+\mathscr{T}_w)=\mathcal{S}/\mathcal{S}_{lam}$ is the total wall shear stress, relative to the laminar unforced value (see \eqref{eq-wss}), and the superscript ``$A$'' indicates the forced case (no superscript is used in the unforced case). 
Note that $\mathcal{FR}_{lam/turb}$ does not take into account the extra drag $\mathscr{B}$ (refer to \eqref{eq-NSz}) produced by the baffle. The net power saving is given by

\begin{equation}
\mathcal{PR}_{lam/turb}:=\frac{(1+\beta)_{turb}-(1+\beta)^{A}_{lam/turb}}{(1+\beta)_{turb}}=\frac{\mathcal{I}_{turb} - \mathcal{I}_{lam/turb}^{A}}{\mathcal{I}_{turb}}\,,
\label{DR}
\end{equation}
where $(1+\beta)=\mathcal{I}/\mathcal{I}_{lam}$ is a measure of the total input energy $\mathcal{I}$ needed to drive the flow at a constant mass flux (refer to \eqref{eq-1plusbeta} and \eqref{eq-energy}).
 In the unforced case, using the Blasius approximation \citep{blasius-1913}, $(1+\beta)_{turb} =(1+\mathscr{T}_w)_{turb} = 0.00494375Re^{0.75}$. \Rone{Turbulent quantities are time averaged over a suitable time window $[(T-\tau), T]$ (refer to \eqref{eq:timeavg}) to exclude initial transients (the notation $\overline{(\bullet)}^t$ in \eqref{DRsk} and \eqref{DR} is omitted for the sake of space). Typically, we use a time horizon $T=600$ and $\tau=400$, i.e. we average over $t=[200-600]$. For $Re=15\,000$, the latter window is shifted by a hundred time units. }
 As noticed earlier for $Re=3000$, typically, $\mathcal{PR} <0$ immediately downstream of the baffle.
However, in the relaminarised cases, assuming the flow stays laminar, a net power saving ($\mathcal{PR} > 0$) will be achieved at a critical downstream distance $z>L_{even}$ where the stress reduction at the wall compensates for the extra drag produced by the baffle.

\begin{figure}
  \centering
      \includegraphics[width=0.5\textwidth]{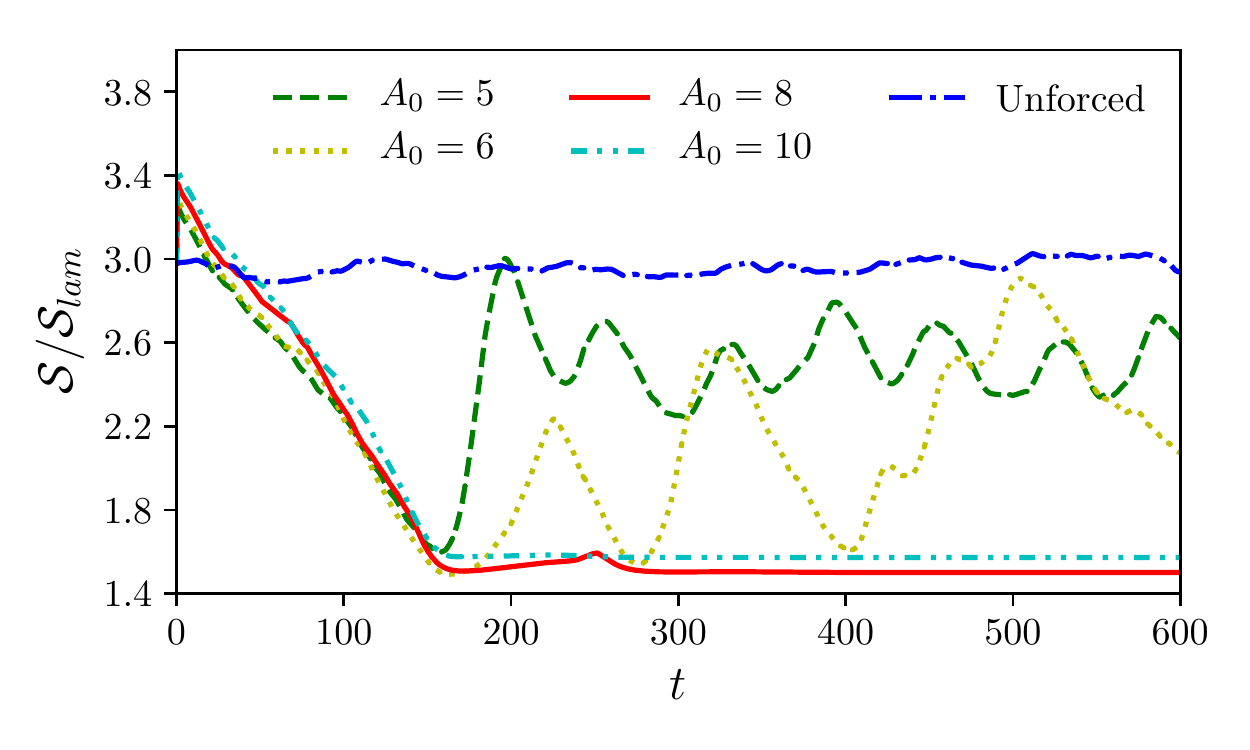}
      \includegraphics[width=0.5\textwidth]{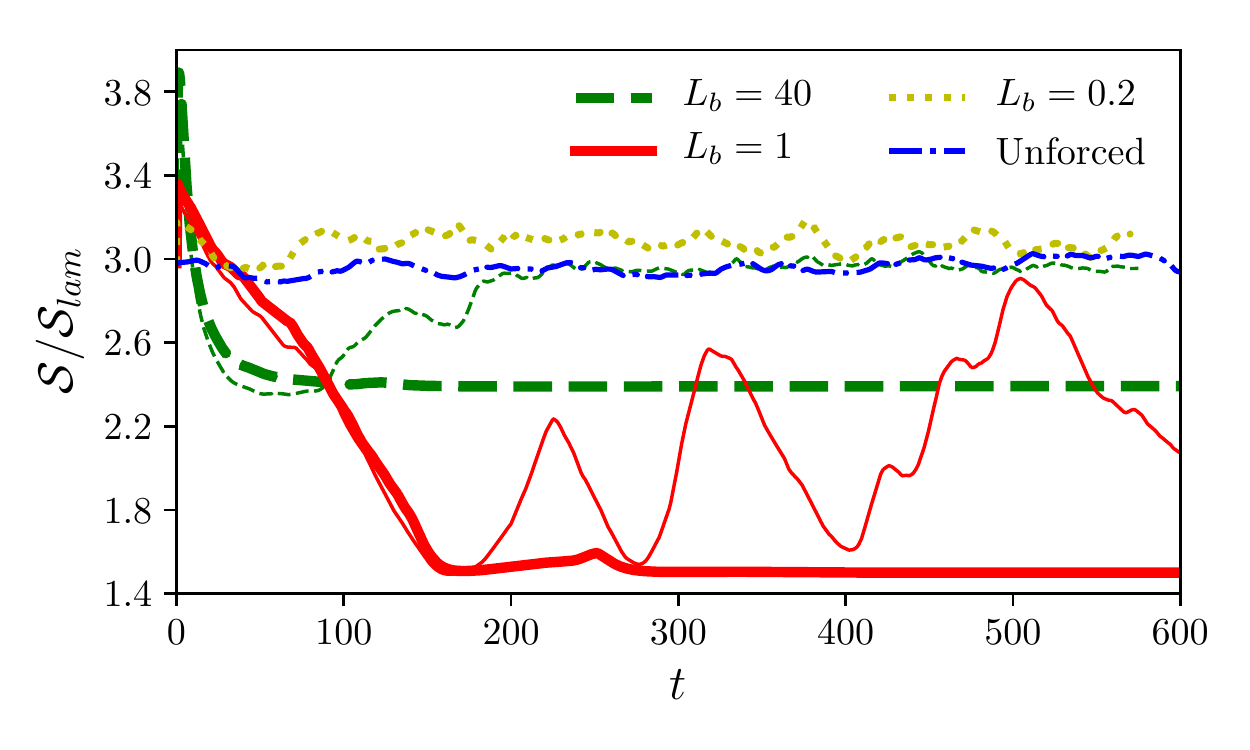}
     \caption{Time series of wall shear stress at $Re=5000$ in the case $L=50$, using the optimal baffle found in \S \ref{sec:result_pb1} for $Re=3000$. Left: effect of $A_0$ for fixed $L_b=1$ (the spatial distribution of $\chi(r,z)$ corresponds to that shown in figure \ref{optF-longpipe}). Right: effect of $L_b$ for amplitudes $A_0$ that can (thick lines) and cannot (thin lines) relaminarise the flow, \Rthree{namely, for $L_b=40$ (green dashed line), $A_0=10$ (thin line) and $A_0=12$ (tick line); for $L_b=1$ (red solid line), $A_0=6$ (thin line) and $A_0=8$ (tick line); for $L_b=0.2$, $A_0=10$.}}
\label{fig-re5000}
\end{figure}

First, at $Re=5000$, we studied the effect of $A_0$ on a typical turbulent field for a fixed $L_b=1$. Starting from the minimal value $A_{crit}=2.9$ obtained at $Re=3000$, we gradually increased $A_0$ until relaminarisation was achieved.
As shown in figure \ref{fig-re5000} (left), an amplitude $A_0 \ge 8$ is needed in order to relaminarise the flow, almost three times larger than at $Re=3000$. For values of $A_{0}\ge 8$, a considerable reduction of the wall shear stress is achieved with respect to the unforced case. As previously observed for $Re=3000$ (results not shown), even when $A_0$ is not large enough to relaminarise the flow, we are still able to 
reduce the wall shear stress considerably (see, e.g., the case $A_0=6$)
and the flow exhibits an interesting periodic time behaviour (see, e.g., the case $A_0=5$), which may suggest
\Rone{that a periodic orbit is approached.}
\Rthree{The shear stress reduction in the `partially' relaminarised cases and the approach to a simpler dynamics, characteristic of lower Reynolds number flows,
 are in line with the numerical results of \citet{kuhnen-etal-2018b}.}
At $Re=5000$ we also verified that the 
streamwise localised baffle ($L_b=1$) is still the `optimal', i.e. it is able to 
 reduce the wall shear stress more than the longer baffle ($L_b=40$), as shown in figure \ref{fig-re5000} (right). Furthermore, the long baffle cannot produce any skin-friction turbulent drag reduction when it cannot relaminarise the flow. 
 As for $Re=3000$, relaminarisation is not possible when $L_b< L_{b,min} =1$ and in this case the wall shear stress is higher than in the unforced case (see the case $L_b=0.2$ in figure \ref{fig-re5000}). Therefore, we fix $L_b=1$.

\begin{figure}
  \centering
      \includegraphics[width=0.495\textwidth]{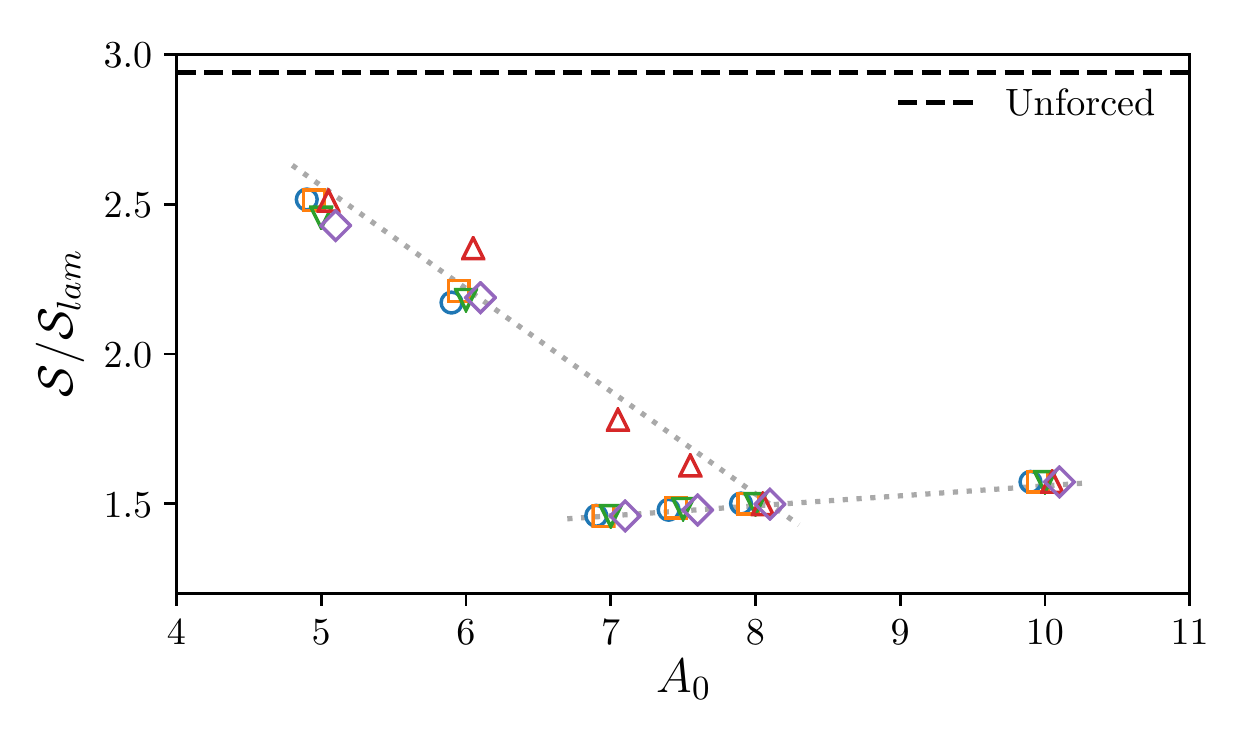}
      \includegraphics[width=0.495\textwidth]{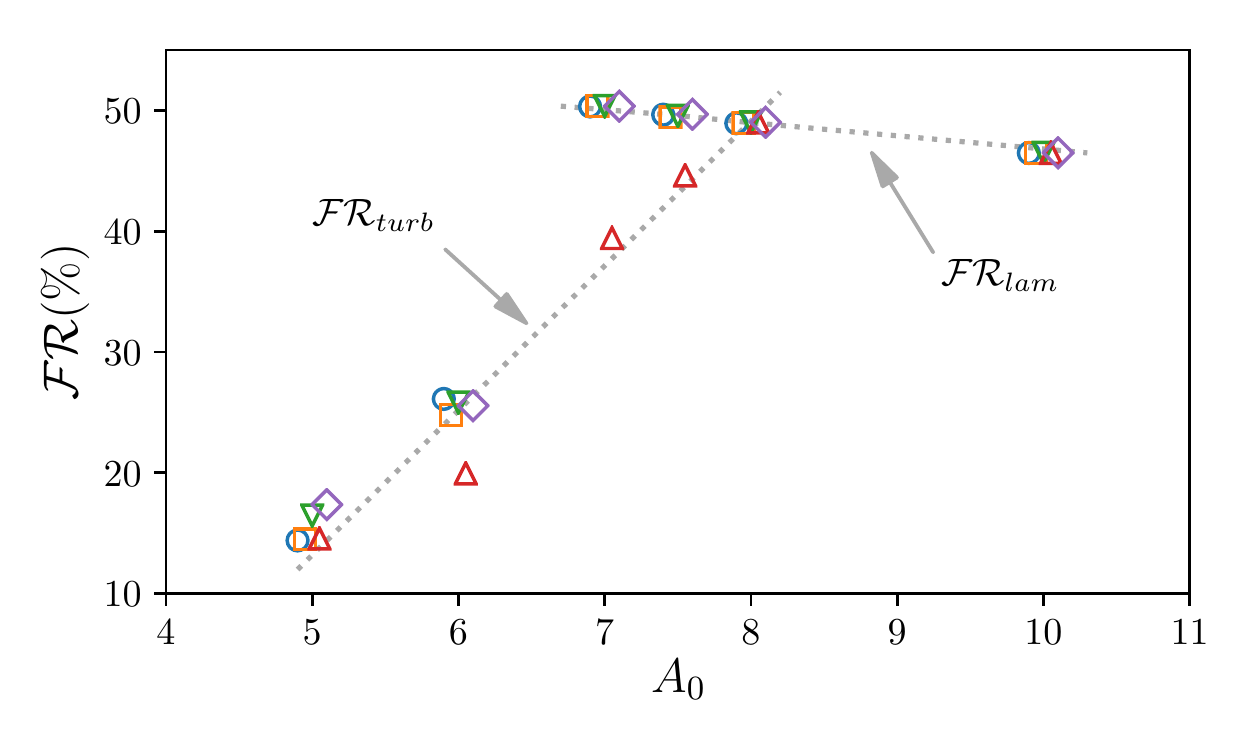}
     \caption{Wall shear stress and skin-friction drag reduction vs $A_0$ for 5 turbulent initial conditions (indicated with different symbols) at $Re=5000$. The black horizontal dashed line in the left graph indicates the Blasius approximation for a turbulent unforced flow, while the dotted grey lines are drawn to guide the eye.}
\label{fig-fric-re5000}
\end{figure}

The effect of the baffle amplitude $A_0$ at $Re=5000$ on the wall shear stress $\mathcal{S}/\mathcal{S}_{lam}$, and on the corresponding skin-friction drag reduction $\mathcal{FR}$, is also seen in figure \ref{fig-fric-re5000} for the 5 turbulent initial fields considered at this Reynolds number.
The critical amplitude $A_{crit}=8$ that can relaminarise all 5 initial conditions divides the curve of $\mathcal{FR}$ (or $\mathcal{S}/\mathcal{S}_{lam}$) vs $A_0$ in two branches:
 for $A_0>A_{crit}$ we have skin-friction laminar drag reduction (all initial conditions lie on the upper branch of $\mathcal{FR}$ corresponding to $\mathcal{FR}_{lam}$), while for $A_0<A_{crit}$ skin-friction turbulent drag reduction is obtained (at least one initial condition lies on the lower branch of $\mathcal{FR}$, corresponding to $\mathcal{FR}_{turb}$). As $A_0$ approaches $A_{crit}$, more initial conditions move to the upper (laminar) branch. Note, however, that at this critical amplitude $A_{crit}=8$ for relaminarisation, a break-even distance $L_{even}$ from the baffle of almost $500$ is needed to achieve a net power saving, as discussed later (see the curve for $Re=5000$ in figure \ref{fig-re-effect-DR}).

\begin{figure}
  \centering
      \includegraphics[width=0.495\textwidth]{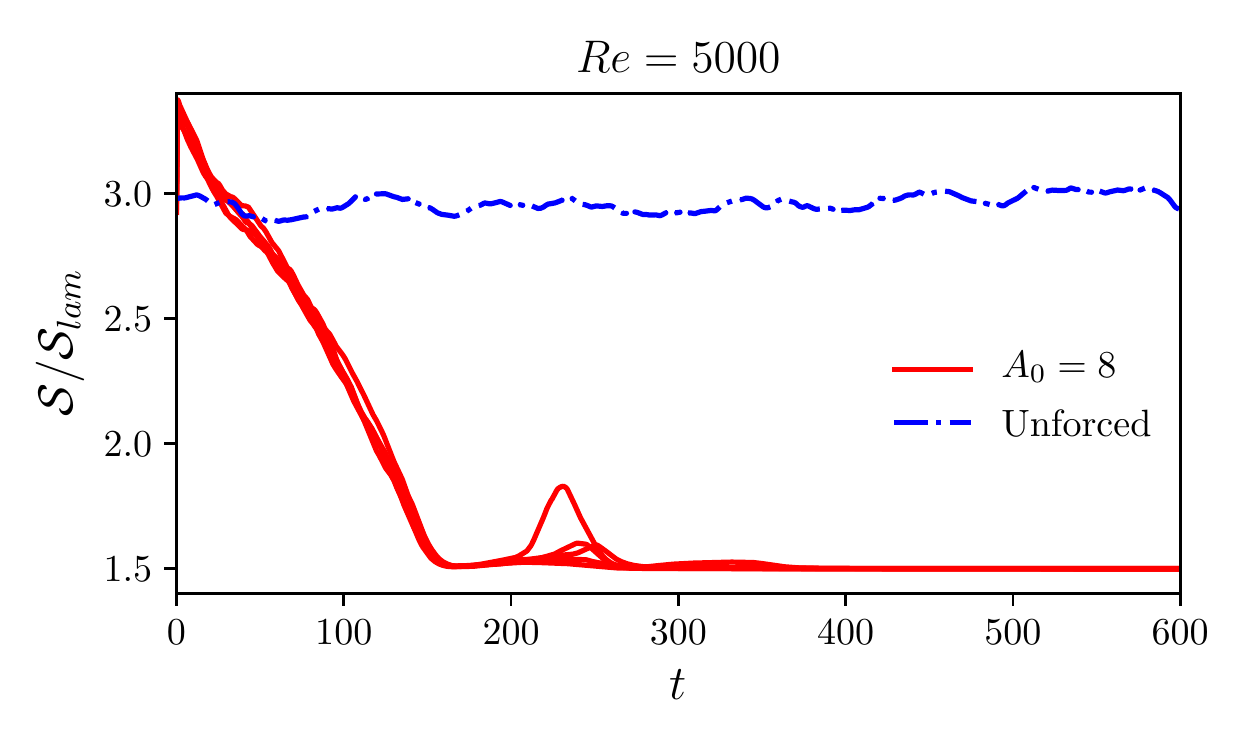}
      \includegraphics[width=0.495\textwidth]{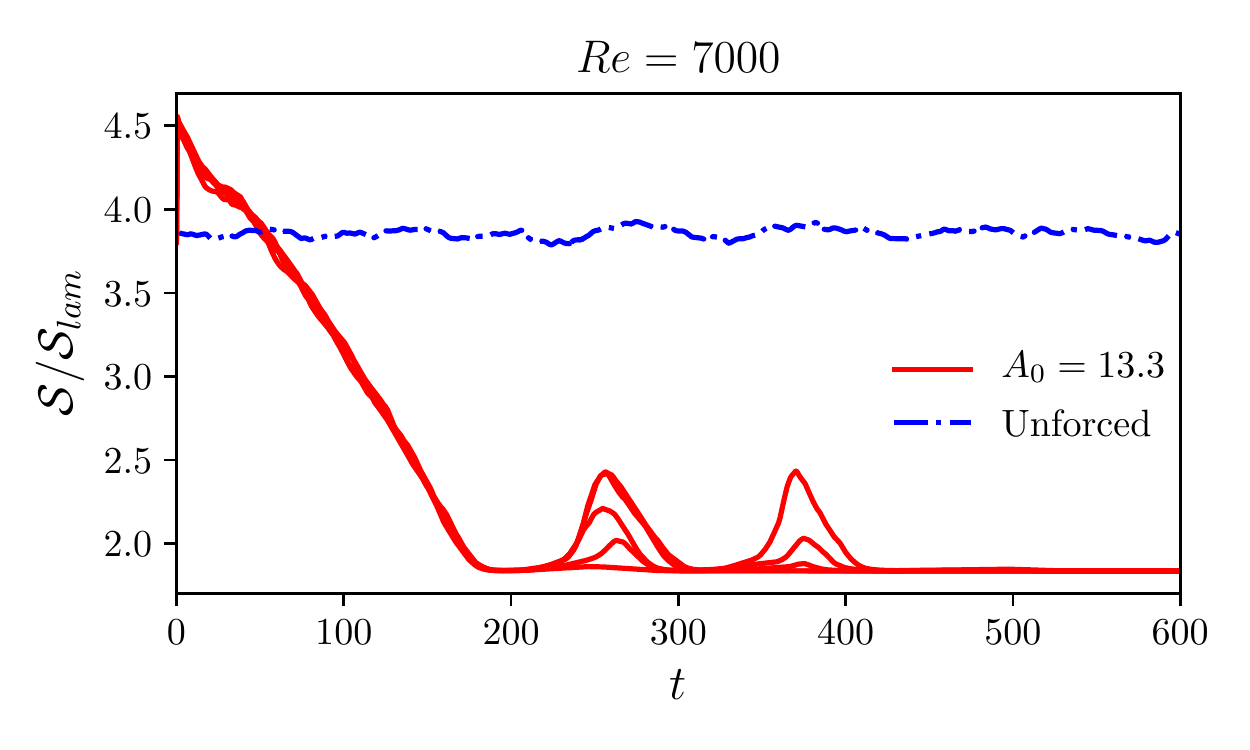}
      \includegraphics[width=0.495\textwidth]{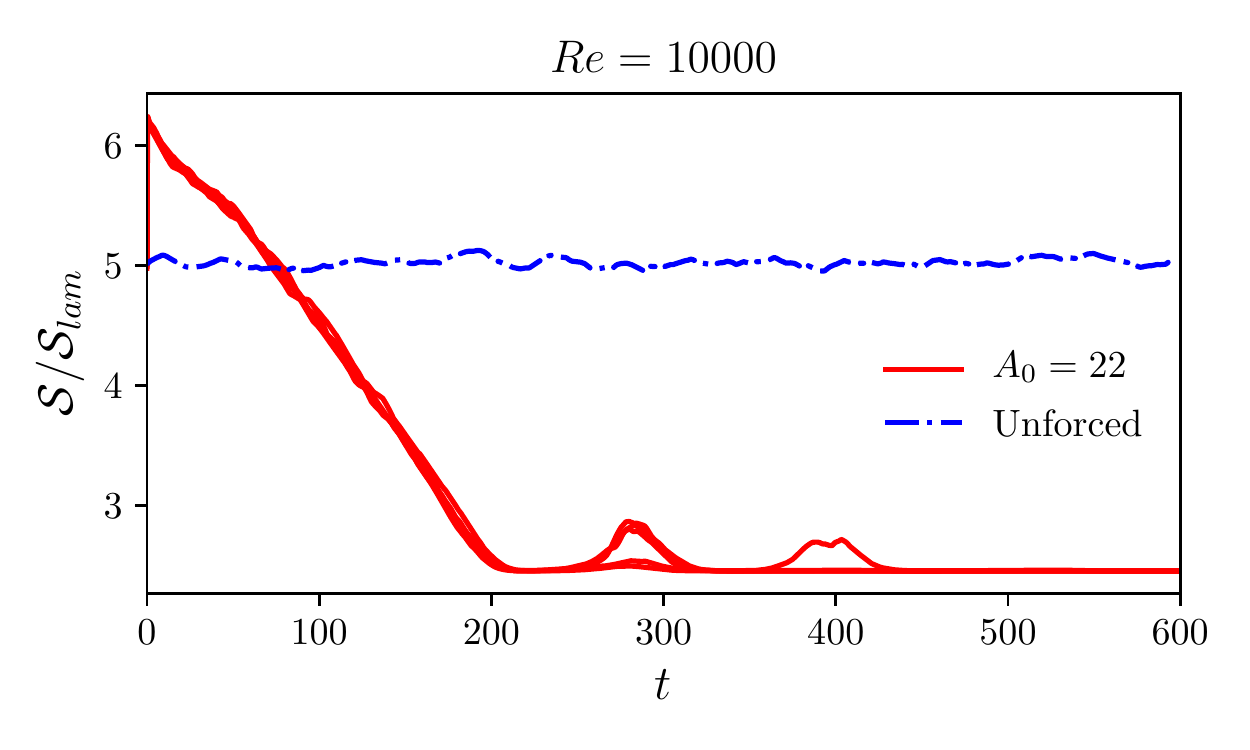}
      \includegraphics[width=0.495\textwidth]{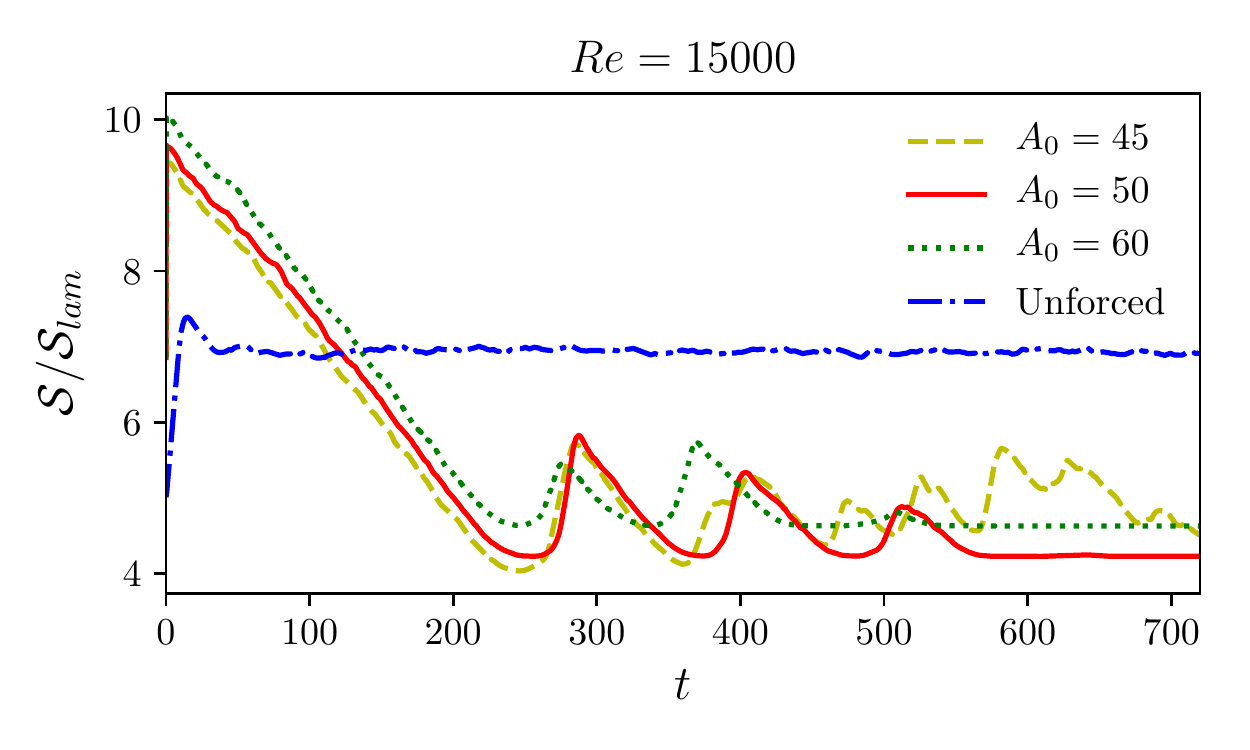}
     \caption{Time series of wall shear stress for different Reynolds numbers in the case $L=50$. The spatial distribution of $\chi(r,z)$ is fixed as the one shown in figure \ref{optF-longpipe} and $N=5$ turbulent fields are used as initial conditions at each Reynolds number (except at $Re=15\,000$, for which only one turbulent initial condition was used to limit the computational cost).}
\label{fig-re-effect}
\end{figure}

\begin{figure}
  \centering
      \includegraphics[width=0.495\textwidth]{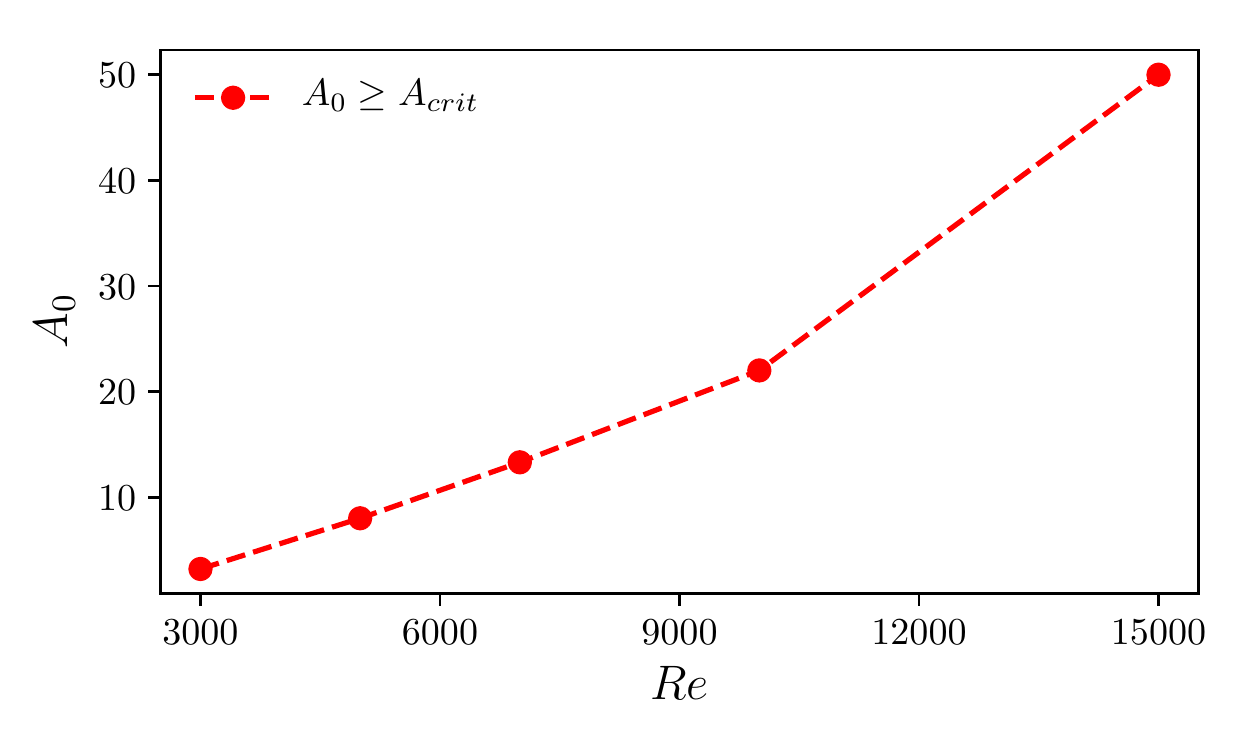}
      \includegraphics[width=0.495\textwidth]{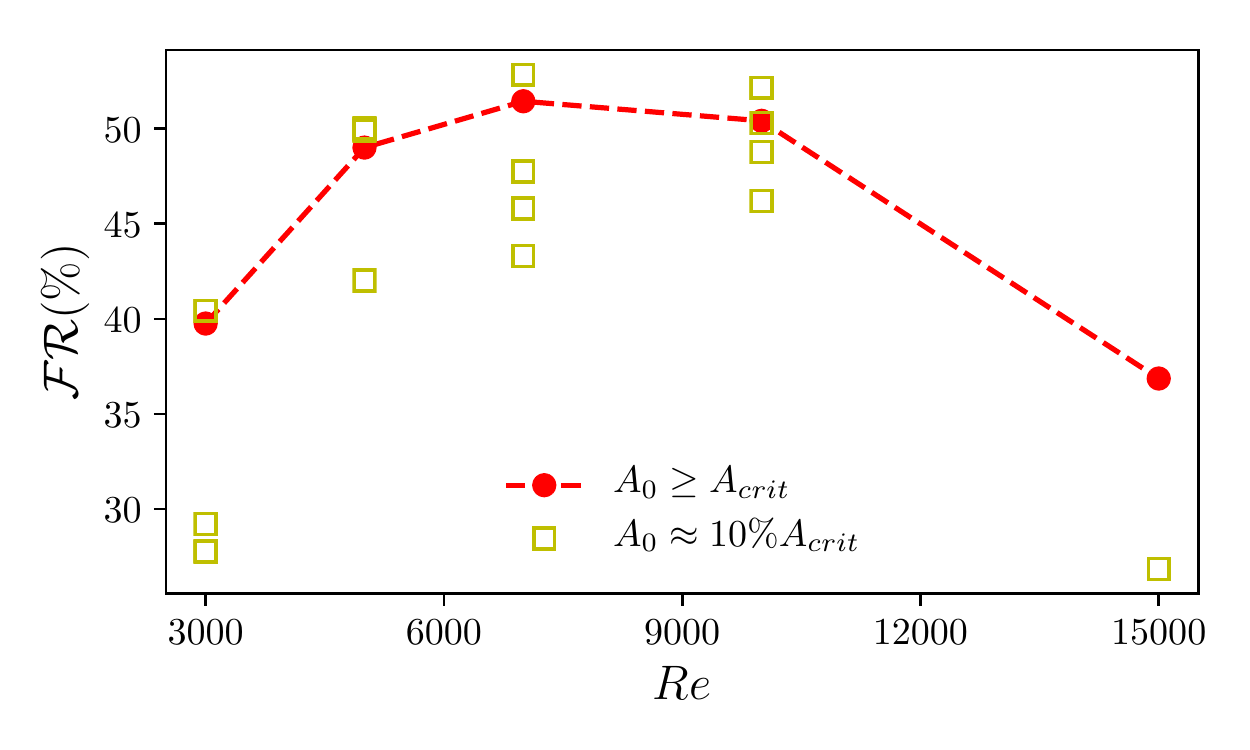}
     \caption{Left: Baffle drag amplitude $A_0$ just above (can relaminarise \emph{all} given turbulent initial fields) the critical value $A_{crit}$ as a function of the Reynolds number $Re$. Right: the corresponding skin-friction laminar drag reduction (red dashed line with filled circles). The skin-friction turbulent drag reduction obtained for $A_0=10 \% A_{crit}$ is also shown with symbols (yellow hollow squares) for the five initial conditions considered at each $Re$. Note, for example, that at $Re=5000$, for $A_0=10 \% A_{crit} \approx 7$, some initial conditions have relaminarised already, as it was also shown in figure \ref{fig-fric-re5000}. The situation is analogous for the other Reynolds numbers.}
\label{fig-re-effect-FR}
\end{figure}

\begin{figure}
  \centering
      \includegraphics[width=0.6\textwidth]{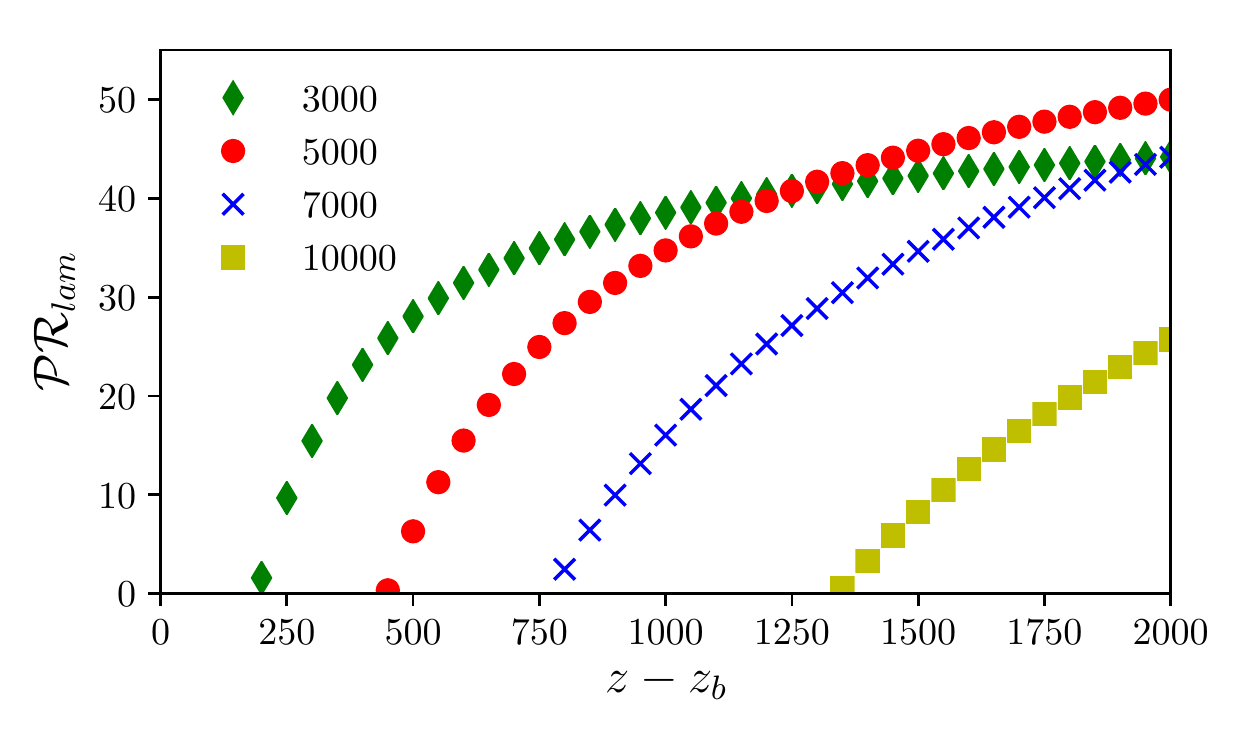}
     \caption{Net power saving as a function of the downstream distance $z$ from the baffle (centred at $z=z_b$) for different $Re$ at $A_0\ge A_{crit}$. The intersections of the curves with the $x-$axis ($\mathcal{PR}=0$) correspond to the critical downstream distances $L_{even}$ needed to achieve a net power saving at different $Re$.}
\label{fig-re-effect-DR}
\end{figure}

The same procedure -- i.e. gradually increase the amplitude of the optimal baffle until all turbulent initial conditions relaminarise -- was carried out for the other Reynolds numbers. Figure \ref{fig-re-effect} shows that after appropriate rescaling of $A_0$ a full collapse of turbulence is obtained up to $Re=15\,000$. The curves of $\mathcal{S}/{S}_{lam}$ and $\mathcal{FR}$ vs $A_0$ for $Re=7000 - 15\,000$ are analogous to those shown in figure \ref{fig-fric-re5000} for $Re=5000$.
Figure \ref{fig-re-effect-FR}(left) shows how the amplitude for relaminarisation varies with the Reynolds number. When $Re$ is increased from $3000$ to $15\,000$, $A_0$ needs to be increased by almost an order of magnitude in order to obtain relaminarisation. Both the laminar and turbulent skin-friction drag reductions also increase with $Re$ up to $Re=7000 - 10\,000$, as shown in the right graph of figure \ref{fig-re-effect-FR}.
However, the increase in $A_0$ with $Re$ is also accompanied by a (roughly linear) increase in the critical length $L_{even}$ for a net energy saving, as shown in figure \ref{fig-re-effect-DR}. At $Re=3000$ the break-even distance $L_{even}$ is a modest $200$, but it reaches $\approx 2700$ at $Re=15\,000$ (not shown). 
Nevertheless, for a downstream distance $z \ge L_{even}$, a net power saving is achieved and, for example at $Re=3000$, $\mathcal{PR}=20\%$ for $z-z_b \approx 300$.

\section{Conclusions}
\label{sec:conclusions}
Motivated by recent experimental studies \citep{kuhnen-etal-2018a} of forced relaminarisation by a stationary obstacle, we have tackled the problem of designing the optimal baffle to destabilise the turbulence in pipe flows. The drag force exerted by the baffle was modelled in our simulations as $\fvec(\mathbf{x},t)$=$-\chi(\mathbf{x})\mathbf{u}_{tot}(\mathbf{x},t)$, where $\mathbf{u}_{tot}$ is the total (laminar flow plus perturbation) velocity field and $\chi(\mathbf{x})\ge 0$ is a measure of the ``level of blockage" of the flow by the baffle. An optimisation algorithm was developed and numerically solved at $Re=3000$ in order to find the optimal spatial distribution of $\chi(\mathbf{x})$, characterised by a given amplitude $\langle \chi(\mathbf{x})\rangle=A_0$, where the angle brackets indicate volume integral. The variational problem was formulated as a minimisation problem for the total viscous dissipation $\mathcal{D}(\mathbf{u}_{tot})$, averaged over a sufficiently long time horizon $T$, subject to the above amplitude constraint on $\chi$, as well as the constraints of the three-dimensional continuity and Navier-Stokes equations and constant mass flux. The amplitude $A_0$ was then gradually decreased until the critical value $A_{crit}$ for relaminarisation was reached, below which turbulence could not be suppressed. The alternative, more efficient, optimisation problem of minimising directly the total energy input $\mathcal{I}(\mathbf{u}_{tot}; \chi) = \mathcal{D}(\mathbf{u}_{tot}) + \mathcal{W}(\mathbf{u}_{tot}; \chi)$, where $\mathcal{W}(\mathbf{u}_{tot}; \chi) = \langle -\fvec \cdot \uvec_{tot} \rangle$ is the work done by the flow against the baffle drag, was also investigated. With such choice of objective functional, the amplitude constraint on $\chi$ is not needed as the algorithm is allowed to vary (typically decrease) $A_0$ in order to deliver the smallest, most energy efficient, baffle. The numerical solution of this problem, however, suffered from convergence issues as baffles that do not relaminarise the flow could be encountered, thus preventing convergence due to sensitivity to initial conditions. Therefore, the latter optimisation problem, once fed with very good initial guesses for $\chi$, was primarily used to confirm the outcomes of the first optimisation problem. Improvements to the formulation and numerical solution of the second optimisation problem, such as adding a constraint on the $L_1$ norm (or a different norm) of $\chi$, will be considered in our future research, possibly for optimising different types of body forces, as discussed in the last paragraph of this section.

Two pipe lengths were considered: $L=10$ and $L=50$. In both cases, starting from $N>1$ (typically $N=20$) turbulent velocity fields and suitable initial guesses for $\chi(\mathbf{x})$, the algorithm converged to an optimal shape of $\chi$ characterised by a strong radial concentration close to the wall and symmetry around the pipe axis, but was found to be slow in organising the streamwise structure. 
Therefore, the optimal streamwise extent of the baffle $L_b$ was sought manually by fitting an analytical radial profile to the optimal one found by the optimisation algorithm and performing direct numerical simulations with different $L_b$. The wall shear stresses and the input energies were monitored in order to quantify the benefit of the baffle. In the long-pipe case $L=50$, a streamwise localised baffle was found to reduce the shear stress at the pipe wall and the input energy more than a wide baffle, with a minimum $L_b$ of approximately one radius, below which relaminarisation was not possible.

Therefore, we fixed the shape of the baffle to be the optimal one found at $Re=3000$ in a $L=50$ pipe and studied the effect of the Reynolds number on the performance of the baffle by means of DNS. We considered the Reynolds number range $Re=5000-15\,000$ and the simulations were fed with $N=5$ different turbulent initial conditions at each $Re$. After suitable rescaling of the amplitude, the optimised baffle, was found to fully relaminarise the flow up to $Re=15\,000$ ($Re_{\tau}\approx 450$). Large shear-stress reductions were found at the wall with a maximum skin-friction drag reduction of more than 50\% at $Re = 7000$ ($Re_{\tau}\approx 230$). The baffle, however, introduces an extra drag due to the local pressure drop. A sufficiently long section of laminar flow downstream of the baffle is thus needed in order to achieve a net energy gain, i.e. where the stress reduction at the wall compensates for the extra drag caused by the baffle. 
At $Re=3000$ the break-even distance $L_{even}$ is a modest $200$, consistent with the experiments of \cite{kuhnen-etal-2018a}, and at a distance $z-z_b\approx300$ from the baffle a $20\%$ net power saving is achieved. However, due to the large amplitude of the baffle at large $Re$, $L_{even}$ increases (approximately linearly) with the Reynolds number and reaches $L_{even} \approx 2700$ at the largest Reynolds number, $Re=15\,000$, considered here.

Our results showed that this purely passive method can relaminarise the flow up to a relatively high Reynolds number, but is not very energy efficient. Other, more general, types of body forces $\fvec=\boldsymbol{\varphi}(\mathbf{x})$ or $\fvec=\boldsymbol{\varphi}(\mathbf{x},t)$, with an active component (i.e. $\boldsymbol{\varphi}$ does not have to oppose the flow everywhere in the domain) might lead to a better performance and increased energy savings. 
\Rthree{A useful starting point may be the volume force used by \citet{kuhnen-etal-2018b}.}
Optimising such types of body forces is the focus of our future research.
\section{Acknowledgements}
This work was funded by EPSRC grants EP/P000959/1 (EM \& APW) and EP/P001130/1 (ZD \& RRK). Fruitful discussions with Bj{\"o}rn Hof and Davide Scarselli are kindly acknowledged. EM would like to thank Dr Luca Sortino for help with figure 1. The authors would like to thank the anonymous referees for their useful suggestions and comments.

\appendix


\section{Optimisation using $L_2$ norm}
\label{app_A}
The formulation using $L_2$ norm is presented in the following. The Lagrangian becomes:
\beq
\lagr_1= \mathcal{J}_1 +  \lambda \left[\langle \chi^2(\mathbf{x}) \rangle - A_0\right] + \sum_n \int_0^T \left \langle \vvec_n \cdot \left[\mathbf{NS} (\uvec_n) + \chi(\mathbf{x})\mathbf{u}_{tot,n}(\mathbf{x},t) \right]\right \rangle \mathrm{d}t + ...
\label{lagrangian-L2}
\eeq
The gradient and the update for the next iteration are:
\beq
\frac{\delta\lagr_1}{\delta \chi} = 2 \lambda \chi(\mathbf{x}) + \sigma(\mathbf{x}) =0\,,
\label{dLdchi-L2}
\eeq
\beq
\chi^{(j+1)}=\chi^{(j)} - \epsilon  \frac{\delta\lagr_1}{\delta \chi^{(j)}} = \chi^{(j)} -\epsilon \left[ 2 \lambda \chi^{(j)}(\mathbf{x}) + \sigma^{(j)}(\mathbf{x}) \right]\,,
\label{update-L2}
\eeq
where $\sigma(\mathbf{x})$ is defined in \eqref{eq-sigma}. All the rest is unchanged. To ensure the update is non-negative, \eqref{update-L2} is replaced by:
\beq
\chi^{(j+1)} = \max\left(0,\,\chi^{(j)}-\epsilon\left(2\lambda \chi^{(j)}+\sigma^{(j)}\right)\right)\,.
\label{update-L2-positive}
\eeq
To find $\lambda$, we impose that $\left \langle \left[\chi^{(j+1)}(\mathbf{x})\right]^{2} \right \rangle =A_0$ and we employ a bracketing method (e.g. the regula falsi algorithm) to find the root $\lambda$ of $\mathcal{H}(\lambda)=\left \langle \left[\chi^{(j+1)}(\mathbf{x})\right]^{2} \right \rangle -A_0 = \left \langle \left[\max\left(0,\chi^{(j)}-\epsilon\left(2\lambda \chi^{(j)}+\sigma^{(j)}\right)\right)\right]^{2} \right \rangle - A_0=0$.

\begin{figure}
  \centering
      \includegraphics[width=1\textwidth]{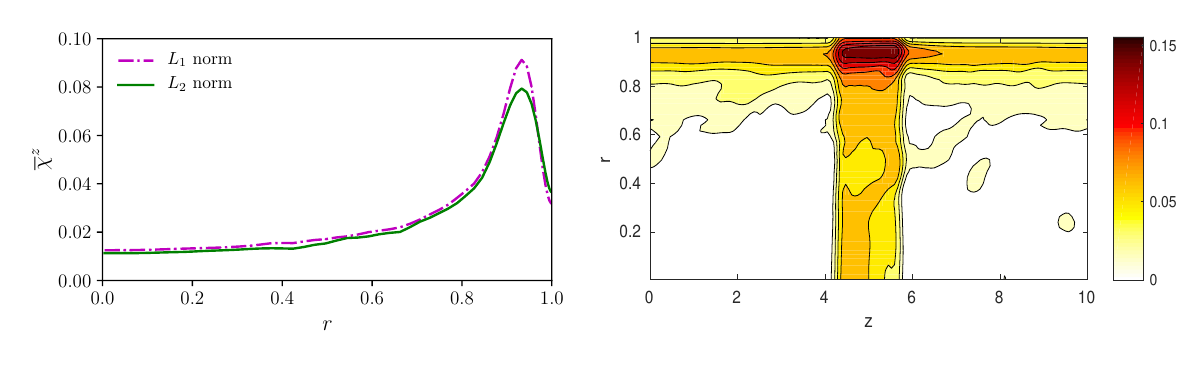}
     \caption{Case $L=10$, $T=300$, with $\mathcal{B}_3$ as initial guess (right plot of figure \ref{fig-verythin}). Comparison of the optimal baffle obtained using $L_1$ and $L_2$ norms. Left: optimal radial profile $\overline{\chi}^z(r)$. The $L_1$-normed distribution (purple dash-dotted curve) is the same as that shown in figure \ref{fig-cfr-opt-radprof} with the same colour/line style. Right: $r-z$ cross section of $\chi$ obtained with the $L_2$-norm constraint.}
\label{fig-cfr-chi00}
\end{figure}

Figure \ref{fig-cfr-chi00} shows a comparison of the optimal baffle obtained using $L_1$ and $L_2$ norms starting from the initial guess shown in figure \ref{fig-verythin}(left). The optimal radial profiles (left) are similar, both presenting the radial concentration close to the wall. The one obtained using the $L_2$ norm is slightly less peaked, as it is reasonable to expect. The formulation with $L_2$ norm is also found to be weakly dependent on the $z-$support, as shown in the $r-z$ cross section on the right.
\section{Formulation for optimisation problem 2}
\label{app_B}
By taking variations of the Lagrangian given in \eqref{lagrangian-opt2} and setting them equal to zero we obtain the following set of Euler-Lagrange equations:\\

\emph{Adjoint Navier-Stokes and continuity equations}
\begin{subequations}
\begin{align}
\frac{\delta \lagr_2}{\delta \uvec_n} &= \frac{\p \vvec_n}{\p t} + \mathcal{U}\frac{\p \vvec_n}{\p z} -\mathcal{U}'\,v_{z,n} \hat{\mathbf{r}} + \nabla \times (\vvec_n \times \mathbf{u}_n) -  \vvec_n \times \nabla \times \mathbf{u}_n +\nabla \Pi_n \,+ \\ \noindent \nonumber & +\frac{1}{Re} \nabla^2\vvec_n  -\Gamma_n(t)\hat{\mathbf{z}} - \phi^2(\mathbf{x})\vvec_n+\frac{2}{Re \, T}\nabla^2\uvec_{tot,n} -\frac{2}{T}\phi^2\uvec_{tot,n}  =0\,,\\
 \frac{\delta \lagr_2}{\delta p_n} &=  \nabla \cdot \vvec_n=0\,.
\end{align}
\label{adj-NS-2}
\end{subequations}


\emph{Compatibility condition}
\beq
\frac{\delta \lagr_2}{\delta \uvec_n(\mathbf{x},T)}=\vvec_n(\mathbf{x},T)= 0\,.
\label{compat-2}
\eeq

\emph{Optimality condition}
\beq
\frac{\delta\lagr_2}{\delta \phi} =\phi\tilde{\sigma}(\mathbf{x}) =0\,,
\label{optim-2}
\eeq
where
\beq
\tilde{\sigma}(\mathbf{x}) :=\sum_n \int_0^T \uvec_{tot,n}\cdot\left(\frac{\uvec_{tot,n}}{T} +\vvec_n\right)\mathrm{d}t
\label{sigma-tilde}
\eeq
 is a scalar function of space. 
Note that the adjoint Navier-Stokes equations \eqref{adj-NS-2} are the same as \eqref{adj-NS} for optimisation problem 1 with the additional forcing term $ -2\phi^2\uvec_{tot,n}/T $. 
The update for the next iteration is
\beq
\phi^{(j+1)}=\phi^{(j)} - \epsilon  \frac{\delta\lagr_2}{\delta \phi^{(j)}} = \phi^{(j)} -\epsilon \phi^{(j)} \tilde{\sigma}^{(j)}(\mathbf{x}) = \phi^{(j)}[1-\epsilon\tilde{\sigma}^{(j)}(\mathbf{x})].
\label{update-2}
\eeq
Note that with this optimisation problem we still have the issue that if $\phi=0$ initially, it cannot change.\\

\section{Spectral filtering}
\label{app_C}
\subsection{Optimisation problem 1}
A spectral filtering on $\phi$ is implemented by adding a constraint to the Lagrangian, namely
\beq
\lagr_1= \sum_n\overline{\mathcal{D}_n}^t(\uvec_n) +\lambda \left[\langle \phi^2(\mathbf{x}) \rangle - A_0\right] + \xi \left \langle \phi -\mathcal{F}(\phi)\right \rangle +...,
\label{lagr-filter}
\eeq
where $\mathcal{F}$ is the spectral filtering operator.
 Using the linearity of $\mathcal{F}$, it is straightforward to show that the above filtering constraint is equivalent to applying the filter to the gradient $\delta \lagr_1/\delta \phi$, that is
\beq
\mathcal{F}\left(\frac{\delta\lagr_1}{\delta \phi}\right) = \mathcal{F}\left( \phi\left(\lambda + \sigma(\mathbf{x})\right) \right) =0,
\eeq
where $\sigma$ is defined in \eqref{eq-sigma}. The update is thus:
\beq
\phi^{(j+1)}=\phi^{(j)} - \epsilon \mathcal{F}\left( \phi^{(j)}\left(\lambda + \sigma^{(j)}(\mathbf{x})\right) \right) =  \phi^{(j)} - \epsilon \lambda \phi^{(j)} - \epsilon \mathcal{F}\left(\sigma^{(j)} \phi^{(j)} \right),
\eeq
where $\mathcal{F}\left(\phi^{(j)}\right)= \phi^{(j)}$. In order to find $\lambda$ we impose
\beq
 \left \langle \left[\phi^{(j+1)}\right]^2 \right \rangle = \left \langle \left[\phi^{(j)}-\epsilon \mathcal{F}\left(\sigma^{(j)} \phi^{(j)}\right) -\phi^{(j)} \epsilon \lambda\right]^2 \right \rangle= A_0.
\eeq

\subsection{Optimisation problem 2}

With the filter $\phi=\mathcal{F}(\phi)$ the optimality condition becomes:
\beq
\mathcal{F}\left(\frac{\delta \lagr_2}{\delta \phi}\right) = \mathcal{F}\left(\phi \tilde{\sigma}\right)=0\,,
\eeq
where $\tilde{\sigma}$ is defined in \eqref{sigma-tilde}. The update is:
\beq
\phi^{(j+1)} =  \phi^{(j)} - \epsilon  \mathcal{F}\left(\frac{\delta \mathcal{L}}{\delta \phi^{(j)}}\right) =\phi^{(j)} -\epsilon\mathcal{F}\left(\phi^{(j)}\tilde{\sigma}^{(j)}\right) = \mathcal{F}(\phi^{(j+1)}).
\eeq

\section{The $L_1$ amplitude condition} 
\label{app_D}

To understand the significance of the set of $\boldsymbol{x}$ which maximize $-\sigma( \boldsymbol{x} )$, we generalise the $L_1$ amplitude constraint used in the main body of the paper to 
\beq
\| \chi\|_\alpha:=\left(   \frac{1}{V} \int \chi^\alpha dV   \right)^{1/\alpha} = \frac{A_0}{V} = a\,,
\label{D1}
\eeq
where $V$ is the volume of the pipe, so that limit $\alpha \rightarrow 1$ recovers the $L_1$ condition. The modified Lagrangian becomes
\begin{align}
\lagr_1 = &\ldots + \lambda \left[ V \left( \frac{1}{V}\int \chi^\alpha \mathrm{d}V   \right)^{1/\alpha}   - A_0\right] +  \sum_n\int_0^T \left \langle \vvec_n \cdot \left[\ldots+ \chi (\mathbf{x})\mathbf{u}_{tot,n}(\mathbf{x},t) \right]\right \rangle \mathrm{d}t  
+\ldots \nonumber \\
& -\int \mu( \boldsymbol{x} ) (\chi - \gamma^2) \mathrm{d}V\,,
\label{D2}
\end{align}
where the requirement that $\chi$ is positive semidefinite is now explicitly imposed.
Setting variations with respect to $\chi$, $\mu$ and $\gamma$ to zero gives, respectively,
\begin{align}
\lambda \,\alpha \,a^{1-\alpha} \chi^{\alpha-1}+\sigma-\mu &=0, \label{D3}\\
\chi-\gamma^2 &=0, \label{D4} \\
2 \mu \gamma & =0.  \label{D5}
\end{align}
Equation (\ref{D5}) makes it clear that either: i) $\gamma=0$ (so $\chi=0$ and $\sigma=\mu$); or ii) $\mu=0$; or iii) both, if $\sigma$ happens to vanish.  
The baffle therefore can only exist in places where $\gamma(\boldsymbol{x}) \neq 0$ which requires $\mu(\boldsymbol{x})=0$ (although the opposite is not true) and, in this case, the relations (\ref{D1}), (\ref{D3})-(\ref{D5}) simplify to just (\ref{D1}) and (\ref{D3}) rearranged as follows
\begin{align}
\left(   \frac{1}{V} \int \left[ \frac{\chi}{a} \right]^\alpha \mathrm{d}V   \right)^{1/\alpha} & = 1 \label{D6}\,, \\
\left[ \frac{\chi}{a} \right]^{\alpha-1}&= \frac{-\sigma}{\alpha \lambda}\,. \label{D7}
\end{align}
Substituting (\ref{D7}) into (\ref{D6}) gives
\beq
\label{D8}
\alpha \lambda = \|  -\sigma \|_{1/\eps}:=\biggl[  \frac{1}{V} \int (-\sigma)^{1/\eps} \mathrm{d}V  \biggr]^{\eps}\,,
\eeq
where $\eps:=(\alpha-1)/\alpha$ goes to $0^+$ as $\alpha \rightarrow 1^+$, and then
\beq
 \frac{-\sigma}{\alpha \lambda} =\frac{-\sigma}{\|  -\sigma \|_{1/\eps}}.
\eeq
%
%
%
%
At this point, it is worth stressing that $\sigma=\sigma(\boldsymbol{x};\eps)$ and, assuming the simple regular expansion
\beq
\sigma(\boldsymbol{x};\eps)=\sigma_0(\boldsymbol{x})+\eps\sigma_1(\boldsymbol{x})+O(\eps^2)\,,
\label{D10}
\eeq
where $\sigma_0(\boldsymbol{x})$ is the distribution calculated in \S \ref{subsec:res_IG}, then
\beq
\label{D11}
(-\sigma)^{1/\eps}=(-\sigma_0(\boldsymbol{x}) )^{1/\eps} e^{\sigma_1(\boldsymbol{x})/\sigma_0(\boldsymbol{x})+O(\eps)} \,.
\eeq
In the limit $\eps \rightarrow 0$ ($\alpha \rightarrow 1$), substituting \eqref{D11} into \eqref{D8}, it follows that, $\lambda \rightarrow \max_{\boldsymbol{x}} (-\sigma_0(\boldsymbol{x}) )$ and so $-\sigma_0(\boldsymbol{x})/\lambda \leq 1$, as discussed in \S \ref{subsec:res_IG}.
For the baffle structure, there are two scenarios - either $\sigma_0$ achieves its global maximum at isolated points or there are finite domains over which this is achieved. Consider the former first. Equation (\ref{D7}) is 
\beq
\left[ \frac{\chi}{a} \right]^\alpha = \frac{ (-\sigma)^{1/\eps} }{ \frac{1}{V} \int (-\sigma)^{1/\eps} \mathrm{d}V } \,.
\label{D12}
\eeq
If $-\sigma_0^*=\max_{\boldsymbol{x}} (-\sigma_0)$ occurs at an isolated point $\boldsymbol{x}^*$, then
\beq
\frac{1}{V} \int (-\sigma)^{1/\eps} \mathrm{d}V
\approx 
c \eps^{d/2} e^{\sigma_1(\boldsymbol{x}^*)/\sigma_0^*} (-\sigma_0^*)^{1/\eps} 
\label{D13}
\eeq
using Laplace's method in $d$ dimensions and letting $c$ gather all the known constants together. 
So, using \eqref{D11} and \eqref{D13} in \eqref{D12}, we obtain
\beq
\left[ \frac{\chi}{a}  \right]^\alpha = 
\frac{1}{ c \eps^{d/2} }
e^{   \sigma_1 (\boldsymbol{x})/\sigma_0(\boldsymbol{x}) - \sigma_1 (\boldsymbol{x}^*)/\sigma_0^*  }
\biggl[ \frac{ -\sigma_0(\boldsymbol{x}) }{ -\sigma_0^* } \biggr]^{1/\eps}
\label{D14}
\eeq
which, as $\eps \rightarrow 0$ ($\alpha \rightarrow 1$), either tends to zero if $\boldsymbol{x} \neq \boldsymbol{x}^*$ or diverges to infinity at $\boldsymbol{x} =\boldsymbol{x}^*$ in such a way that the volume integral is finite. Hence the baffle structure is a $\delta$-function or a collection of $\delta$-functions around the set of isolated $\boldsymbol{x}^*$ which globally maximize $-\sigma_0(\boldsymbol{x}) $.

The alternate, and more plausible, scenario is that $-\sigma_0$ is maximized over a connected set $\Lambda:=\{\boldsymbol{x} \,| -\sigma_0(\boldsymbol{x})=\max_{\boldsymbol{x}}(-\sigma_0)\}$ (or sets). Then (\ref{D14}) is replaced by 
\beq
\left[ \frac{\chi}{a}  \right]^\alpha \approx \frac{e^{\sigma_1(\boldsymbol{x})/\sigma_0(\boldsymbol{x})}}
                                                                   { \int_{\Lambda}e^{\sigma_1(\boldsymbol{x})/\sigma_0^*} \mathrm{d}V }
\biggl[ \frac{ -\sigma_0(\boldsymbol{x}) }{ -\sigma_0^* } \biggr]^{1/\eps}
\label{D15}
\eeq
as $\eps \rightarrow 0$. Assuming $\mu=0$ over $\Lambda$, the baffle then has a smooth structure over $\Lambda$ generated by $\sigma_1(\boldsymbol{x})$ ({\em a posteriori} justification for adopting the simple regular expansion in (\ref{D10})\,). However, there can be subsets of $\Lambda$ over which $\mu \neq 0$ where the baffle vanishes. This possibility allows for a non-uniqueness of the optimal baffle which can only exist on the intersection of the set where $\mu=0$ and $\Lambda$. 

\bibliographystyle{jfm}
\bibliography{./pipes}

\end{document}